\documentclass[spanish,english]{article}
\usepackage[T1]{fontenc}
\usepackage[latin9]{inputenc}
\usepackage[a4paper]{geometry}
\usepackage{amsmath}
\usepackage{amsthm}
\usepackage{amssymb}
\usepackage{cancel}
\usepackage{esint}
\usepackage[all]{xy}

\makeatletter
\usepackage{babel}

\makeatother

\usepackage{babel}
\addto\shorthandsspanish{\spanishdeactivate{~<>}}

\begin{document}
\title{The Spacetime Picture in Quantum Gravity}
\author{{\normalsize{}{}Alejandro Ascárate}}
\maketitle
\begin{abstract}
{\normalsize{}{}We propose an approach which, by combining insights
from Loop Quantum Gravity (LQG), Topos theory, Non-commutative Geometry}\emph{\normalsize{}{}
}à\emph{\normalsize{}  la}{\normalsize{}{} Connes, and spacetime
relationalism, provides fertile ground for the search of an adequate
spacetime picture in Quantum Gravity. With this approach, we obtain
a novel way of deducing the quantization of the possible values for
the area of a surface. One gets the same area values than those from
the area operator in standard LQG, but our approach makes a further
prediction: some smaller values and sub-divisions are also allowed.
In addition, the }\textit{\normalsize{}{}area}{\normalsize{}{} arises
as a noncommutative }\emph{\normalsize{}{}distance}{\normalsize{}{}
between two noncommutative points, and thus they should be interpreted
as irreducible string-like objects at the physical level (where the
area interpretation for the noncommutative distance holds).}{\normalsize\par}

{\normalsize{}{}$\;$}{\normalsize\par}

\end{abstract}

\subsection*{{\normalsize{}{}Introduction}}

$\;$

One of the reasons why General Relativity (GR) is so compelling, and
its formalism so \emph{intuitive} to handle (once we get used to the
revolutionary changes it makes to our intuitive, pre-relativistic
notions of space and time), is that the theory is based on a \emph{spacetime
picture formulation}. By this we mean that the main object of the
theory is the pair 
\[
(M,g_{ab}),
\]
where $M$ is a $4-$dimensional differentiable manifold (with the
standard topological requirements) and $g_{ab}$ a smooth lorentzian
metric on it \cite{key-1}. While the manifold $M$ represents \emph{physical
spacetime,} the metric $g_{ab}$ allows us to calculate quantities
which represent such things as physical distances, times, and volumes
in it. Space and time are physical notions which unlike (say) quarks,
we \emph{directly experience}, and this is what gives GR its intuitive
appeal. This is also useful since it tends to be easier to develop
a theory when we have a clear picture of what it describes at the
physical level.

$\;$

In the case of Quantum Gravity (QG), however, it's not at all clear
what the thing that this theory supposedly describes actually is.
Of course, we actually know that it should describe the quantum behavior
of gravity and spacetime geometry, since GR describes the classical
behaviour of these notions, and QG is supposedly a quantization of
the first. But, so far (and leaving aside the obvious fact that actually
we don't even have a working theory of QG) no \emph{clear} picture
of these quantum phenomena is available, that is, one which can have
an appeal and \emph{physico-mathematical role} similar to that of
the spacetime picture of classical GR, i.e. we lack a picture of \emph{quantum
spacetime}.

$\;$

That doesn't mean that we don't have candidates or proposals for a
QG theory ---we have plenty of them. And what makes things even more
unclear is that the various proposals look very different from each
other. The proposals of interest here will be three\footnote{Actually, only i) is a theory of QG (in the sense of quantizing GR):
ii) and iii) are abstract mathematical theories whose aim is to study
geometrical properties of non-commutative spaces. They are sometimes
mentioned as QG candidates because it's assumed that their techniques
will be useful in the mathematical formulation of QG (indeed, that's
one of the motivations of the authors of those theories). We, of course,
agree with that view, and actually our aim here will be to put it
into practice.}:

$\;$

i) Loop Quantum Gravity (LQG) \cite{key-2,key-6};

$\;$

ii) Non-commutative Geometry à\emph{  la} Connes (NCG) \cite{key-4};

$\;$

iii) the Topos Approach to Quantum Theory \cite{key-3}.

$\;$

The task is to find an analogue in QG of the spacetime pair $(M,g_{ab})$
of classical GR. But to do this, we need to carefuly study what exactly
spacetime is at the deep physical level, in order to know how to proceed.
The key that allows us to reach a \emph{convergence} of these approaches
is the philosophical view in which spacetime emerges from matter in
a \emph{relational} way \cite{key-5,key-6}. For the analogue of the
spatial manifold $\varSigma\subset M$, for example, we will propose
a framework in which ``\emph{quantum $3-$dimensional space}'' is
the \emph{spectral presheaf} $\underline{\varSigma}$ (a concept taken
from iii)) based on the quantum kinematical algebra of Algebraic Quantum
Gravity (AQG), which is an alternative to LQG (albeit closely related
to it) that doesn't assume a background differentiable manifold and
has also a better behaved classical limit.

$\;$

\subsection*{{\normalsize{}{}1. Algebraic Reformulation of Classical Space(-time)}}

$\;$

We will only deal with physical $3-$space for now, since time requires
more elaborated considerations. We assume $\mathcal{A}$ to be a \emph{commutative}
$C^{*}-$algebra (with unit). We know that, by using the \emph{Gelfand
transform}, there's a compact topological space $\varSigma$ such
that $\mathcal{\mathcal{A}}$ gets represented as the algebra $C(\varSigma)$
of complex-valued continuous functions on $\varSigma$ (note that
applying the transform to $C(\varSigma)$ itself will give this very
same algebra). Thus, if $\mathcal{\mathcal{A}}$ was the space algebra,
then we get the space picture, i.e. the space $\Sigma$, by this process.
Our aim here will be to try to replicate this process in some way
for the quantum case. That is, we take the space algebra $\mathcal{\mathcal{A}}$
as the central object and the space picture as something secondary
or derived from it. More precisely, the space $\varSigma$ is given,
in the commutative case, by the Gelfand spectrum, that is, the set
of algebraic states $\omega:\mathcal{\mathcal{A}}\,\longrightarrow\mathbb{C}$
which are also algebraic homomorphisms between $\mathcal{\mathcal{A}}$
and $\mathbb{C}$; it turns out that these also comprise the set of
all possible pure states. Thus, we want to study the space algebra
$\mathcal{\mathcal{A}}$ and its pure states $\omega$, which represent
the space ``points''. In the standard approach, the structure of
$\varSigma$ is simply assumed to be that of a compact topological
space, which in turns leads to a commutative structure for the algebra.
Here we focus only on the algebra $\mathcal{\mathcal{A}}$ and its
pure states $\omega,$ where the set of the latter only gives a compact
topological space $\varSigma$ if $\mathcal{\mathcal{A}}$ is commutative,
which is precisely what we \textit{won't} take for granted.

$\;$

Now, both in classical GR and in LQG, a commutative structure is given
to the algebra of physical $3-$space. Of course, this is well suited
for the former, but it seems problematic for the latter, since geometric
properties like area are discretized. Thus, if we follow that approach,
we risk giving space an a priori structure and also risk tangling
into contradictions between assumptions and conclusions. We don't
want to assume any a priori structure on space, in particular about
the commutativity of the space algebra. Thus, we adopt a ``space
algebra first'' approach rather than the usual ``space points first''
one. This is something natural to do if we adopt the so-called algebraic
approach to quantum theories, in which one reconstructs the standard
formalism of quantum theory from an abstract quantum ``phase space''
algebra \cite{key-7}. Recall from what we said that one can introduce
a structure on the ``points space'' and obtain from it the structure
of the algebra of functions, or one can introduce a structure on the
latter and from there try to deduce the one of the points space, in
particular by taking the pure states on the algebra as the points;
in the commutative case, the Gelfand transform links the two approaches
and makes them equivalent, but if the space algebra is non-commutative
(as we suspect), then the manifold disappears (and, obviously, the
Gelfand equivalence too) and therefore a space algebra first approach
is more appropriate, since the algebra is what survives even in the
non-commutative case.

$\;$

$\mathbf{Definition\;1.1}$: The (kinematical) phase space of GR is
defined as

\[
X=\left\{ \left[h_{ab},\pi^{ab}\right]\diagup h_{ab},\pi^{ab}\in C^{\infty}(\varSigma)\,(\mathrm{as\,fields}\,\mathrm{on}\,\varSigma)\right\} ,
\]
where $h_{ab}$ and $\pi^{ab}$ are, respectively, a smooth riemannian
metric on a spacelike Cauchy hypersurface $\Sigma$ in a compact and
boundaryless spacetime $M$, foliated by $\Sigma$ as usual, and the
conjugate momentum tensor density$.\,\blacksquare$

$\;$

$\mathbf{Definition\;1.2}$:\footnote{We will not discuss the topology of the phase space here, but the
functionals defined here should be continuous under any reasonable
topology on it, since the determinant of $h_{ab}$ is just a linear
combination of products of its components.} The subset $\mathcal{F}\subset C(X)$ consists of the phase space
functionals of the form

\[
F_{f}\left(\left[h_{ab},\pi^{ab}\right]\right)\doteq\int_{\varSigma}f\boldsymbol{\epsilon}(h_{ab}),\,\forall\left[h_{ab},\pi^{ab}\right]\in X,
\]
where $f\in C^{\infty}(\varSigma)$ and $\boldsymbol{\epsilon}(h_{ab})$
is the volume element of $h_{ab}$, i.e. $\boldsymbol{\epsilon}(h_{ab})=\sqrt{h}\,\mathrm{d}^{3}x$
($h=\mathrm{det}\,(h_{ab})$)$.\,\blacksquare$

$\;$

$\mathbf{Proposition\;1.1}$: The assignment $f\,\longmapsto F_{f}$
is injective.

$\;$

\emph{Proof}: since the phase space point $\left[h_{ab},\pi^{ab}\right]$
is composed by arbitrary smooth functions, $\sqrt{h}$ behaves as
an arbitrary smooth function when one varies $\left[h_{ab},\pi^{ab}\right]$
over \emph{all} of phase space, and this makes the assignment $f\,\longmapsto F_{f}$
injective, since 
\[
\int_{\mathcal{\varSigma}}f\,\sqrt{h}\,\mathrm{d}^{3}x=\int_{\mathcal{\varSigma}}f'\,\sqrt{h}\,\mathrm{d}^{3}x,\,\forall h\in C^{\infty}(\Sigma)
\]
\[
\implies\int_{\mathcal{\varSigma}}(f-f')\,\sqrt{h}\,\mathrm{d}^{3}x=0,\,\forall h\in C^{\infty}(\Sigma)
\]
\[
\implies f-f'=0\text{ almost everywhere}
\]
so by continuity $f-f'$ vanishes identically i.e. $f=f'$.$\,\square$

$\;$

$\mathbf{Definition\;1.3}$: We now define a mapping $R:\mathcal{F}\times X\,\longrightarrow C^{\infty}(\varSigma)\times Obj(\mathsf{Hil})$
(where $Obj(\mathsf{Hil})$ is the collection of objects in the category
$\mathsf{Hil}$ of Hilbert spaces) by 
\[
(F_{f},\left[h_{ab},\pi^{ab}\right])\,\longmapsto R(F_{f},\left[h_{ab},\pi^{ab}\right])\doteq(f,L^{2}(\mathcal{S},\boldsymbol{\epsilon}(h_{ab})),
\]
(where $\mathcal{S}$ is the $C^{\infty}(\varSigma)-$module of smooth
spinor fields in $\varSigma$, which from now on we assume allows
a spin structure)$.\,\blacksquare$

$\;$

$\mathbf{Definition\;1.4}$: For \emph{fixed} $\left[h_{ab},\pi^{ab}\right]\in X$
and \emph{variable} $F_{f}\in\mathcal{F}$, we denote the first component
of $R(F_{f},\left[h_{ab},\pi^{ab}\right])$ as $R_{h}(F_{f}).\,\blacksquare$

$\;$

$\mathbf{Definition\;1.5}$: We make $\mathcal{F}$ into an (unital)
algebra $(\mathcal{F},$$\cdot_{sp.})$, where the product $\cdot_{sp.}$
in $\mathcal{F}$ is defined as

\[
\left(F_{f_{1}}\cdot_{sp}F_{f_{2}}\right)\left(\left[h_{ab},\pi^{ab}\right]\right)\doteq F_{f_{1}f_{2}}\left(\left[h_{ab},\pi^{ab}\right]\right),\,\forall\left[h_{ab},\pi^{ab}\right]\in X
\]

\[
=\int_{\varSigma}f_{1}f_{2}\boldsymbol{\epsilon}(h_{ab}).\,\blacksquare
\]

$\;$

All of the previous constructions were introduced in order to see
the algebra $C^{\infty}(\varSigma)$ as coming from a subset of phase
space functionals once a point in $X$ is taken. Indeed, $(\mathcal{F},\cdot_{sp})$
is carried by $R_{h}$ to the pointwise product of space functions
(i.e. the product of $C^{\infty}(\varSigma)$). We want this because
the process of canonical quantization only gives us the quantum phase
space algebra, so we must adapt everything to this and formulate our
concepts accordingly.

$\;$

$\mathbf{Corollary\;1.1}$: The map $R_{h}$ is a \emph{bijection}
beetwen $\mathcal{F}$ and $C^{\infty}(\varSigma)$ (by Proposition
1.1), and is a faithfull algebra representation of $\mathcal{F}$
into\footnote{$\mathcal{B}(\mathcal{H})$ is the algebra of all the \emph{bounded}
operators that act on the Hilbert space $\mathcal{H}$, with the operator
composition $\circ$ as algebraic product.} $\mathcal{B}\left(L^{2}(\mathcal{S},\boldsymbol{\epsilon}(h_{ab}))\right)$
by multiplication operators, i.e. $R_{h}(F_{f})(\psi)(x)\doteq f(x)\psi(x),\,\forall\psi\in L^{2}(\mathcal{S},\boldsymbol{\epsilon}(h_{ab}))$$.\,\square$

$\;$

We call $R_{h}$ the ``\emph{relational representation}'', since
\emph{it gives a way of obtaining the algebra of physical $3-$space
purely from Field properties} (represented here by the phase space
functionals). This is precisely the essence of the spacetime relationist
viewpoint, according to which \textit{points of the spacetime manifold
don't have an existence of their own, but rather acquire physical
meaning only in the context of a particular fixed solution}, where
they can be identified in terms of the values the different fields
take on them in that particular solution (in other words: only once
we fix a solution to the gravitational field equations i.e. a metric,
can we label events with physically meaningful coordinates, such as
distances and proper times). Since we live in a particular solution
and not in phase space (which is more of a conceptual construct of
ours), there is no problem with this, and \textit{the dependency of
our ``relational representation'' on a fixed metric is thus completely
expected.}

$\;$

\emph{Switching to a relational and algebraic frame of mind, we can
take this representation as the actual way of defining how to build
space from the phase space algebra of the gravitational field.}

$\;$

Before continuing, we recall some notions from NCG \cite{key-4}.

$\;$

$\mathbf{Definition\;1.5}$: In a Hilbert space $\mathcal{H}$, 
\begin{enumerate}
\item a \emph{spectral triple} $(\mathcal{A},\mathcal{H},D)$ is a sub $^{*}-$algebra
$\mathcal{A}\subset\mathcal{B}(\mathcal{H})$ together with a self-adjoint
operator $D$ (possibly unbounded, so it's defined on a dense domain
$\mathrm{Dom}\,D$) of compact resolvent (that is, $(D^{2}+1)^{-1}$
is compact) and such that $[D,a]\in\mathcal{B}(\mathcal{H}),\,\forall a\in\mathcal{A}$
(for the commutator to be defined we need, of course, that $a(\mathrm{Dom}\,D)\subseteq\mathrm{Dom}\,D$); 
\item the \emph{purely algebraic distance formula} for a spectral triple
is given by: 
\[
\cancel{d}\,(\varphi,\psi)=\mathrm{sup}\,\left\{ \mid\varphi(a)-\psi(a)\mid,\,\forall a\in\mathcal{A}\diagup\parallel\left[a,D\right]\parallel\leq1\right\} ,
\]
for any two \emph{states} $\varphi,\psi$ (normalized positive linear
functionals on $\mathcal{A}$); 
\item the spectral triple is \emph{even} if there's a self-adjoint unitary
operator $\varGamma$ on $\mathcal{H}$ such $a\Gamma=\Gamma a,\,\forall a\in\mathcal{A}$,
$\varGamma(\mathrm{Dom}\,D)=\mathrm{Dom}\,D$, and $D\Gamma=-\Gamma D$;
if no such operator is given, then the spectral triple is \emph{odd}; 
\item the spectral triple is \emph{real }if there's an antiunitary operator
$J:\mathcal{H}\,\longrightarrow\mathcal{H}$ such that $J(\mathrm{Dom}\,D)\subset\mathrm{Dom}\,D$
and $[a,Jb^{*}J^{-1}]=0,\,\forall a,b\in\mathcal{A}$; it posseses
a \emph{real structure} if, in addition, $J^{2}=\pm1$, $JDJ^{-1}=\pm D$,
and $J\Gamma=\pm\Gamma J$ (in the even case); 
\item a real spectral triple is of \emph{first order} if $[[D,a],Jb^{*}J^{-1}]=0,\,\forall a,b\in\mathcal{A}$; 
\item for $1<p<\infty$, the operator ideal $\mathcal{L}^{p+}(\mathcal{H})$
is defined as\footnote{Where $\mathcal{K}(\mathcal{H})\subset\mathcal{B}(\mathcal{H})$ is
the space of compact operators, and $\sigma_{N}(T)\doteq\sum_{k=0}^{N-1}s_{k}(T)$,
with $s_{k}(T)$ being the $k-$th eigenvalue (in decreasing order
and counted with multiplicity) of the compact positive operator $\mid T\mid\doteq(T^{*}T)^{1/2}$.} 
\[
\mathcal{L}^{p+}(\mathcal{H})\doteq\left\{ T\in\mathcal{K}(\mathcal{H})\diagup\sigma_{N}(T)=O(N^{(p-1)/p}),\,N\,\longrightarrow\infty\right\} ;
\]
if for a \emph{positive} operator $T\in\mathcal{L}^{1+}(\mathcal{H})$
the sequence $\left\{ \frac{\sigma_{N}(T)}{\mathrm{ln}\,N}\right\} _{N\in\mathbb{N}}$
is \emph{convergent}, then its \emph{Dixmier trace} is defined as
\[
\mathrm{tr}^{+}(T)\doteq\underset{_{N\,\longrightarrow\infty}}{lim}\frac{\sigma_{N}(T)}{\mathrm{ln}\,N};
\]
the number $\mathrm{dim}_{Spec.}(\mathcal{A},\mathcal{H},D)=n\in\mathbb{N}$
is the \emph{spectral dimension} of a spectral triple if, for $\mathrm{ker}\,D=\{0\}$,
we have $\mid D\mid^{-1}\in\mathcal{L}^{n+}(\mathcal{H})$ and $0<\mathrm{tr}^{+}(\mid D\mid^{-n})<\infty$;
a spectral triple whose Hilbert space $\mathcal{H}$ is of \emph{finite}
dimension is considered as having spectral dimension $n=0$ (for example,
spectral triples that describe the geometry of a set with a \emph{finite}
number of points, and the Euclidean distances among them, are of this
type); \foreignlanguage{spanish}{the non-commutative integral $\fint$
of algebra elements $a$, is defined by}\footnote{\selectlanguage{spanish}%
See the references for the definition when $a$ is not positive.\selectlanguage{english}%
}\foreignlanguage{spanish}{: 
\[
\fint a\doteq\alpha_{n}\,\mathrm{tr}^{+}(a\mid D\mid^{-n});
\]
} 
\selectlanguage{spanish}%
\item a\foreignlanguage{english}{ spectral triple with a real structure
is of}\footnote{\selectlanguage{english}%
The $KO$ comes from $KO-$homology.\selectlanguage{spanish}%
}\foreignlanguage{english}{ \emph{$KO-$dimension} $\mathrm{dim}_{KO}(\mathcal{A},\mathcal{H},D;J,\varGamma)=2$
($\mathrm{mod}\,8$) if $J^{2}=-1$, $JDJ^{-1}=+D$, and $J\Gamma=-\Gamma J$
(for the rest of possible \emph{$KO-$}dimensions $\mathrm{dim}_{KO}=m\,\mathrm{mod}\,8,\,m\in\mathbb{N}$,
see the table of signs in the cited references)$.\,\blacksquare$} 
\end{enumerate}
$\;$

$\mathbf{Example\;1.1}$: The canonical example of a spectral triple
is the usual commutative geometry of an $n-$dimensional smooth compact
boundaryless $M$ with Riemannian metric $g_{ab}$: $\left(C^{\infty}(M),L^{2}(\mathcal{S},\boldsymbol{\epsilon}(g_{ab})),\cancel{D}_{g}\right)$,
where $C^{\infty}(M)$ acts by multiplication operators, and $\cancel{D}_{g}$
is the \emph{Dirac differential operator}, which, for an orthonormal
basis $\left\{ \boldsymbol{e}_{\alpha}\right\} _{\alpha=1,...,n}$
of the tangent spaces of $M$, is given \emph{locally} by 
\[
\cancel{D}_{g}\psi=-i\sum_{\alpha=1}^{n}\gamma^{\alpha}\nabla_{\boldsymbol{e}_{\alpha}}^{\mathcal{S}}\psi,\,\psi\in\mathcal{S},
\]
where $\nabla_{\boldsymbol{e}_{\alpha}}^{\mathcal{S}}$ is the spin
connection and $\gamma^{\alpha}\in M_{2^{m}}(\mathbb{C})$, $\alpha=1,...,n$
(with $n=2m$ or $n=2m+1$, $m\in\mathbb{N}_{0}$), are the generators
of the action of the Clifford algebra $\mathbb{C}\mathrm{l}(\mathbb{R}^{n})$
on $\mathbb{C}^{2^{m}}$ (and then $\{\gamma^{\alpha},\gamma^{\beta}\}_{M_{2^{m}}(\mathbb{C})}=2\delta^{\alpha\beta}I_{M_{2^{m}}(\mathbb{C})}$,
$\alpha,\beta=1,...,n$); we mention an important result which states
that, if $\triangle^{\mathcal{S}}$ is the spinor Laplacian and $\mathsf{s}$
the scalar curvature, the following formula holds: $\cancel{D}_{g}^{2}=\triangle^{\mathcal{S}}+\frac{1}{4}\mathsf{s}$.
For pure states $\varphi_{p},\varphi_{q},\,p,q\in M$, one gets $\cancel{d}\,(\varphi_{p},\varphi_{q})=d(q,p)$,
where $d$ is the usual distance induced by the metric $g_{ab}$,
and hence $g_{ab}$ is completely characterized by the purely algebraic
and functional analytic information of the spectral triple. One also
gets that $\mathrm{dim}_{Spec.}(\mathcal{A},\mathcal{H},D)=\mathrm{dim}\,M=n$
and that $\fint a_{f}=\int_{M}f\boldsymbol{\epsilon}(g_{ab})$. With
$J$ given by the usual charge conjugation operator and $\varGamma$
given by the $\mathbb{Z}_{2}-$grading of the Clifford algebra, we
get a real structure such that the first order differential operator
$\cancel{D}_{g}$ is also first order in the spectral triple sense,
and such that $\mathrm{dim}_{KO}(\mathcal{A},\mathcal{H},D;J,\varGamma)=\mathrm{dim}\,M\,\mathrm{mod}\,8$.
The celebrated \emph{reconstruction theorem of NCG} states that any
abstract \emph{commutative }spectral triple $(\mathcal{A},\mathcal{H},D)$
that satisfies several regularity assumptions (see references, \cite{key-4}b)
in particular) is always such that $(\mathcal{A},\mathcal{H},D)\cong\left(C^{\infty}(M),L^{2}(\mathcal{S},\boldsymbol{\epsilon}(g_{ab})),\cancel{D}_{g}\right)$
for some \emph{unique} $(M,g_{ab},\mathcal{S})$. A very special sub-case
of commutative geometries are the (spectral) $0-$dimensional triples
corresponding to sets with a \emph{finite} number of points $\{p_{1},...,p_{N}\}$
(say, $2$), and the Euclidean distances among them ($d_{12}$), where
the algebra is given by the \emph{diagonal} $N\times N$ matrices
($2\times2$), with the diagonal corresponding to the values $f(p_{1})$,
$f(p_{2})$, ..., $f(p_{N})$, of functions on the set of points,
and with Dirac operator (in the case of $2$ points) given by $D=\left(\begin{array}{cc}
0 & d_{12}^{-1}\\
d_{12}^{-1} & 0
\end{array}\right)$, which gives $\cancel{d}\,(\varphi_{1},\psi_{2})=d_{12}$; and $\mathrm{tr}^{+}$
can be replaced by the ordinary trace $\mathrm{tr}$ on matrices (thus,
e.g. $\cancel{d}\,(\varphi_{1},\psi_{2})=\fint I\doteq\frac{1}{N}\,\mathrm{tr}\,(\mid D\mid^{-1})=d_{12}$).
Further examples will be given along the way as they are needed$.\,\blacksquare$

$\;$

$\mathbf{Proposition\;1.2}$: If one starts with a subset $\mathcal{A}_{\varSigma}$
(with a commutative product $\cdot_{sp}$) of the phase space algebra
$\mathcal{A}_{X}$ (i.e. $\mathcal{A}_{\varSigma}\subset\mathcal{A}_{X}$,
but only as a set and\emph{ not} as a subalgebra\footnote{This convention will be maintained whenever the symbol $\subset$
is used, unless some other sense is explicitly stated.}), which correspond, respectively, to $\mathcal{F}$ and $C(X)$ but\emph{
seen as abstract algebras}, then there exist a \emph{purely algebraic}
(i.e. which \emph{doesn't }make use of the commutative manifold structure
of space for its definition) faithful representation $\left(\widetilde{R}_{h}\left[\mathcal{A}_{\varSigma}\right],\mathcal{H}_{\widetilde{R}_{h}}\right)$
of $\mathcal{A}_{\varSigma}$ and isomorphisms $\pi_{X}:\mathcal{A}_{\varSigma}\,\longrightarrow\mathcal{F}$
and $\pi_{\varSigma}:\widetilde{R}_{h}\left[\mathcal{A}_{\varSigma}\right]\,\longrightarrow C^{\infty}(\varSigma)$,
such that the following diagram commutes:

\[
\xymatrix{\mathcal{A}_{\varSigma}\ar[d]^{\widetilde{R}_{h}}\ar[r]_{\pi_{X}} & \mathcal{F}\ar[d]_{R_{h}}\\
\widetilde{R}_{h}\left[\mathcal{A}_{\varSigma}\right]\ar[r]^{\pi_{\varSigma}} & C^{\infty}\left(\varSigma\right)
}
\]
i.e. such that

\[
R_{h}=\pi_{\varSigma}\circ\widetilde{R}_{h}\circ\pi_{X}^{-1}.
\]

$\;$

\emph{Proof}: consider a hypothetical faithful representation $\widetilde{R}_{h}$
of $\mathcal{A}_{\varSigma}$ on a Hilbert space $\mathcal{H}_{\widetilde{R}_{h}}$
(i.e. $\widetilde{R}_{h}\left[\mathcal{A}_{\varSigma}\right]\subset\mathcal{B}(\mathcal{H}_{\widetilde{R}_{h}})$,
both as set and as an algebra) and form an \emph{abstract commutative
spectral triple }(satisfying the regularity assumptions) 
\[
\left(\widetilde{R}_{h}\left[\mathcal{A}_{\varSigma}\right],\mathcal{H}_{\widetilde{R}_{h}},D_{h}\right).
\]
Then we can realize those operators as \emph{smooth} functions on
a manifold $\varSigma$ via the Gelfand isomorphsim $\pi_{\varSigma}$
of the reconstruction theorem of NCG, which states that, for a \emph{unique}
metric $h_{ab}$ and spin structure, 
\[
\left(\widetilde{R}_{h}\left[\mathcal{A}_{\varSigma}\right],\mathcal{H}_{\widetilde{R}_{h}},D_{h}\right)\cong\left(C^{\infty}(\varSigma),L^{2}(\mathcal{S},\boldsymbol{\epsilon}(h_{ab})),\cancel{D}_{h}\right),
\]
where $\pi_{\varSigma}\left[\widetilde{R}_{h}\left[\mathcal{A}_{\varSigma}\right]\right]$$=C^{\infty}(\varSigma)$.
Next, one builds the phase space $X$ of Definition 1.1 for this $\varSigma$.
Finally, by the previous uniqueness, we can now recognize that the
resulting $R_{h}$ (for the metric $h_{ab}$) between the \textit{phase
space functions }in $\mathcal{F}$ obtained via the Gelfand isomorphism\footnote{Note that it also acts as an algebra isomorphism in the product $\cdot_{sp}$.}
$\pi_{X}$ (which acts on the whole of $\mathcal{A}_{X}$, i.e. $\pi_{X}\left[\mathcal{A}_{X}\right]=C(X)\supset\mathcal{F}=\pi_{X}\left[\mathcal{A}_{\varSigma}\right]\;$)
on one hand, and\textit{ space functions} on the other, is the concrete
version of the initial $\widetilde{R}_{h}$, i.e. the latter must
be unique, and can therefore simply be defined as $\widetilde{R}_{h}\doteq\pi_{\varSigma}^{-1}\circ R_{h}\circ\pi_{X}$
(with $\pi_{X},\pi_{\varSigma}^{-1}$ defined just using the concrete,
manifold algebras), and the diagram commutes$.\,\square$

$\;$

It's also useful to draw the previous diagram in the following alternative
way, which emphasizes the role of picking a solution $h$ in the construction:

\[
\xymatrix{\mathcal{A}_{\varSigma}\ar[d]^{\widetilde{R}_{h}}\ar[r]_{\pi_{X}} & C(X)\ar[d]_{R_{h}}\\
\mathcal{B}(\mathcal{H}_{\widetilde{R}_{h}})\ar[r]^{\pi_{\varSigma}} & \;\mathcal{B}\left(L^{2}(S_{n},\boldsymbol{\epsilon}(h_{ab}))\right)
}
\]

We understand that the prime advantage of the present approach (where
a representation of a subset of the phase space algebra with certain
products is taken as the definition of the space algebra that relationally
arises from the field, a general construction which obviously applies
to both the classical and quantum cases) is that it provides a \emph{common
abstract procedure} \emph{to build the space algebra in both the classical
and quantum regimes. }Thus, the general process described here is
\emph{free from ad-hoc structures and only uses the structures that
are common to both classical and quantum formalisms, namely, the abstract
phase space algebra and its products}\footnote{but not their specifics, since they can be commutative or non-commutative
and in any combination; in general, the relation between both algebras
is given by the theory in consideration, which can be either classical
field theory or quantum physics. In this sense, what we have exhibited
is only a reconstruction of space and not a general duality between
algebras and theories; this is why in Proposition 1.2 above we assumed
the concrete commutative algebras as a given and only checked for
the overall \emph{consistency} once the relational construction is
applied.}. Note that

$\;$

\emph{in the general, non-commutative case, only the left hand side
of the previous diagram survives}.

$\;$

In this way, \emph{all the geometrical information of the space is
now contained in the spectral triple}.

$\;$

\subsection*{{\normalsize{}{}2. Quantum (or Non-Commutative) Relational Space(-time)}}

$\;$

\subsubsection*{2.1. General Idea}

$\;$

Now, in the case of QG, the phase space algebra is quantized by fusing
its product (given by the pointwise product of phase space functions
and which gives the probability structure of the theory) with the
non-commutative Poisson bracket product, i.e. we get a non-commutative
algebra whose only product has two different physical interpretations
---that's what we mean by ``fusing''. The result is a quantum,
non-commutative (abstract) algebra $\widehat{\mathcal{A}}_{QG}$ which
gives rise to a non-commutative probability theory.

$\;$

$\mathbf{Hypothesis\;2.1}$:\emph{ We make the hypothesis that $\cdot_{sp}$
also gets fused with the other products}\footnote{In this way, it would seem that, in QG, one must go a step further
regarding the scheme of canonical quantization, since we also have
here the deformation of the space product $._{sp}$. It seems legitimate
that this step is only necessary in QG since, in the other quantum
field theories, there's always a background classical gravitational
field, a metric, whose purpose is to give rise to the (classical)
space of the theory (and, since we are at it, to simplify the concepts
and the math), whereas in QG all fields vary and get quantized, and
since in QG a deformed phase space product coexisting with a commutative
$sp$ product is something that could easily lead to internal contradictions,
the most sensible hypothesis seems to be the one we just made.}$.\,\blacksquare$

$\;$

$\mathbf{Definition\;2.1}$: Consider the phase space of GR, $(X,C(X),\left\{ \cdot,\cdot\right\} _{GR})$,
and a background independent Poisson sub-algebra $\mathcal{A}_{GR}$
of phase space functionals there. The basic $^{*}-$algebra, $\widehat{\mathcal{A}}_{QG}$,
in QG will be the algebra $S_{\mathcal{A}_{GR}}$ freely generated
(with field $\mathbb{C}$) by the classical algebra elements\footnote{Which is just the complex vector space generated by the basis $\left\{ e_{S}\right\} $,
where $S$ runs over all the possible ordered, finite sequences of
elements from $\mathcal{A}_{GR}$ (e.g. $S=$$(a_{1},a_{2},...,a_{k}),\,k>0$)
and the algebra product is given at the basis level by $e_{S}\star e_{T}\doteq e_{(S,T)}$.} and with the relations imposed by the ideals $\mathcal{I}_{*LP}$
(where $*$ is the involution, $L$ is from linear and $P$ from $\mathrm{i}$
multiplied by the Poisson brackets. This process simply imposes the
familiar Dirac commutation relations of canonical quantization to
$\widehat{\mathcal{A}}_{QG}$, that is: $\left[\widehat{a},\widehat{a}'\right]_{\widehat{\mathcal{A}}_{QG}}\doteq$$\widehat{a}\star_{\widehat{\mathcal{A}}_{QG}}\widehat{a}'-\widehat{a}'\star_{\widehat{\mathcal{A}}_{QG}}\widehat{a}=$$\mathrm{i}\widehat{\left\{ a,a'\right\} }_{GR}$,
where $\star_{\widehat{\mathcal{A}}_{QG}}$ is the quantum probability
algebraic product of $\widehat{\mathcal{A}}_{QG}$ \cite{key-2}).
That is, $\widehat{\mathcal{A}}_{QG}$ is the quotient:

\[
\widehat{\mathcal{A}}_{QG}\doteq\frac{S_{\mathcal{A}_{GR}}}{\mathcal{I}_{*LP}}.\,\blacksquare
\]
(Note that the subset of classical properties $\mathcal{A}_{GR}$
forms a Lie algebra $\mathfrak{A}_{GR}\doteq(\mathcal{A}_{GR},\left\{ \cdot,\cdot\right\} _{GR})$
with respect to the Poisson bracket, and then the obtained quantum
algebra, by using this quantization process, is its \emph{complex
universal enveloping algebra} $U(\mathfrak{A}_{GR})$, which, unlike
the Lie algebra, is an \emph{associative} algebra. Also note that,
due to the $\mathrm{i}$ in the Poisson brackets, the canonical embedding
map of $\mathfrak{A}_{GR}$ into $U(\mathfrak{A}_{GR})$ is actually
of the form $a\in\mathfrak{A}_{GR}\,\longmapsto(-\mathrm{i}\widehat{a})\in U(\mathfrak{A}_{GR})$.)

$\;$

The above construction defines a \emph{quantization map} $\,\widehat{}:\mathcal{A}_{GR}\,\longrightarrow\widehat{\mathcal{A}}_{QG}$,
$a\,\longmapsto\widehat{a}$, that embeds the elements of the classical
algebra into the quantum one (although, of course, the quantum probability
algebraic relations in the latter are now those determined by the
Poisson bracket and not the pointwise product of $C(X)$; this map
shouldn't be confused with the previous canonical embedding map of
$\mathfrak{A}_{GR}$ into $U(\mathfrak{A}_{GR})$).

$\;$

The next task is to identify a \emph{subalgebra} 
\[
\widehat{\mathcal{A}}_{SP.}\subset\widehat{\mathcal{A}}_{QG}
\]
\emph{that plays a role equivalent to the one of $\mathcal{A}_{\varSigma}$
in the classical case of Proposition 1.2.} The following definitions
are suggested by that proposition and the ``fusing of the products''
assumption of Hypothesis 2.1:

$\;$

$\mathbf{Definition\;2.2}$: Consider the subset $\{f_{S}\}$ of all
elements $f_{S}$ in $\mathcal{F}$ with $S\doteq supp\,f_{S}\subseteq\varSigma$,
then we can see $\mathcal{F}$ as the result of the application of
an \emph{indexation relation} $\widetilde{I}_{\mathcal{F}}:I\,\longrightarrow\mathcal{F},\,S\,\longmapsto\widetilde{I}_{\mathcal{F}}(S)\doteq\{f_{S}\}$,
where $I$ is the collection of all the supports $S$. We now \emph{replace}
$I$ by a subset $I_{Dis}\subset I$ which can be \emph{at most countably
infinite}, and compute the Poisson brackets in $\mathcal{F}_{Dis}\doteq\widetilde{I}_{\mathcal{F}}[I_{Dis}]$:
the subset $I_{Dis}$ should be selected in a way that makes $(\mathcal{F}_{Dis},\left\{ \cdot,\cdot\right\} _{GR})$
a Poisson subalgebra and, in particular, one whose structure constants
\emph{don't }depend on the differentiable manifold details of $\Sigma$
nor on the details of the functions $f_{S}$ varying over it, \emph{but
only on $I_{Dis}$, as an indexing set.} With this set up, we choose
$\mathcal{A}_{GR}$ such that $\mathcal{F}_{Dis}\subset\mathcal{A}_{GR}$
(so that $\widehat{\mathcal{F}}_{Dis}\subset\widehat{\mathcal{A}}_{QG}$)
and define:

\[
\widehat{\mathcal{A}}_{SP.}\doteq\widehat{\mathcal{F}}_{Dis}.\,\blacksquare
\]
(The interpretation of $I_{Dis}$, $\mathcal{F}_{Dis}$ will become
clear in the concrete implementations of the next sections.)

$\;$

Assume for the time being that we have that subalgebra. Then, we need
to study the representation theory of the abstract quantum phase algebra
$\widehat{\mathcal{A}}_{QG}$. Once we have all the representations
\[
\left(R_{QG}\left[\widehat{\mathcal{A}}_{QG}\right],\mathcal{H}_{R_{QG}}\right)
\]
(with $R_{QG}\left[\widehat{\mathcal{A}}_{QG}\right]\subset\mathcal{B}(\mathcal{H}_{R_{QG}})$)
of interest, the representations that will allow us to ``relationally
reconstruct'' the quantum space are simply the \emph{restrictions}
or \emph{subrepresentations} of the previous representations to the
subalgebra $\widehat{\mathcal{A}}_{SP.}\subset\widehat{\mathcal{A}}_{QG}$,
since the phase space product is now the same as the space product.

$\;$

$\mathbf{Definition\;2.3}$: Assuming one has the family of possible
representations $\left(R_{QG}\left[\widehat{\mathcal{A}}_{QG}\right],\mathcal{H}_{R_{QG}}\right)$
of $\widehat{\mathcal{A}}_{QG}$, we define as

\[
\widehat{\mathcal{A}}_{sp.}\doteq R_{QG}\left[\widehat{\mathcal{A}}_{SP.}\right]
\]
the \emph{non-commutative} algebra of quantum physical $3-$space
that ``relationally arises'' from the quantum gravitational field
(this algebra will be the quantum analogue of $\widetilde{R}_{h}\left[\mathcal{A}_{\varSigma}\right]$,
and $R_{QG}$ that of $\widetilde{R}_{h}$, in the left hand side
of the diagram of Proposition 1.2)$.\,\blacksquare$

$\;$

Since we have the quantum space algebra, we now need its ``Gelfand
representation'' to get a \emph{manifold-like} space picture out
of it. Of course, since the algebra is non-commutative, we cannot
apply the standard Gelfand transform to get an ordinary differentiable
manifold space. The Topos approach provides a suitable replacement,
so we now recall some notions from this theory \cite{key-3}.

$\;$

$\mathbf{Definition\;2.4}$: In a Hilbert space $\mathcal{H}$, 
\selectlanguage{spanish}%
\begin{enumerate}
\item the category $\nu(\mathcal{H})$ is defined as the one that has as
objects all commutative von Neumann sub-algebras on $\mathcal{H}$,
and with sub-algebra inclusion as arrows among them; 
\selectlanguage{english}%
\item the \emph{spectral presheaf}, \foreignlanguage{spanish}{$\underline{M}$,}
is the presheaf\footnote{Given a category $C$, a presheaf is a functor $F:C^{\mathrm{op}}\,\longrightarrow\mathsf{Set}$,
the latter being the category of ordinary sets and maps between them.} \foreignlanguage{spanish}{which assigns to an algebra $V\in\nu(\mathcal{H})$
the set $\underline{M}(V)$ given by its Gelfand spectrum, and to
the inclusion $V'\subseteq V$ the map which goes from $\underline{M}(V)\ni\omega:V\,\longrightarrow\mathbb{C}$
to $\underline{M}(V')\ni\omega\mid_{V'}:V'\,\longrightarrow\mathbb{C}$
consisting in reducing the domain of $\omega$ from $V$ to $V'$.
The category made by all presheaves (for a same Hilbert space), with
natural transformations between the functors/presheaves as arrows,
is a Topos} (see the references for the definition of a Topos; informally,
it's a category which ``looks like'' $\mathsf{Set}$ in the sense
that it possesses analogues of \emph{some} of its most relevant features
(such as e.g. a sub-object classifier)$.\,\blacksquare$ 
\end{enumerate}
\selectlanguage{english}%
$\;$

\selectlanguage{spanish}%
The key aspect \cite{key-3} of this construction is that it \emph{bijectively
maps the self-adjoint operators on $\mathcal{H}$ to arrows from $\underline{M}$
to the quantity-value object $\underline{\mathbb{R}}$} (which is
the Topos analogue of the real numbers). In the category of classical
manifolds, the analogue of these arrows are, of course, the functions
from a manifold $M$ to the real numbers $\mathbb{R}$. Thus, the
general \emph{non-commutative} algebra of bounded self-adjoint operators
$\mathrm{A}$ on $\mathcal{H}$ is \emph{mapped} $1-1$ to arrows
$\check{\mathrm{A}}(\mathrm{A}):\underline{M}\,\longrightarrow\underline{\mathbb{R}}$,
in analogy to how the Gelfand isomorphism maps the self-adjoint elements
of a commutative \foreignlanguage{english}{$C^{*}-$algebra (with
unit)} to real valued functions $C(M)\ni f:M\,\longrightarrow\mathbb{R}$
on its Gelfand spectrum $M$ (a compact topological space). This suggest
the following definition:

\selectlanguage{english}%
$\;$

$\mathbf{Definition\;2.5}$: The spectral presheaf $\underline{\varSigma}_{R_{QG}}$
based on $\nu(\mathcal{H}_{R_{QG}})$ is the space picture of quantum
physical $3-$space$.\,\blacksquare$

$\;$

\selectlanguage{spanish}%
Note that, in order to physically interpret this construction as the
analogue of constructing space via the Gelfand transform in the commutative
case, we must first interpret the algebra product in $\widehat{\mathcal{A}}_{sp.}\doteq R_{QG}\left[\widehat{\mathcal{A}}_{SP.}\right]$
as the quantization of the space product (an interpretation in favor
of which we argued in Section 1). Otherwise, even if one has an interpretation
of the elements of the algebra as space ones, the product will still
be the quantized phase space product, and therefore $\underline{\varSigma}$
will be a topos phase space (which is the usual physical interpretation
made in the ``topos approach'' literature) rather than physical
space.

\selectlanguage{english}%
$\;$

$\mathbf{Remark\;2.1}$:\foreignlanguage{spanish}{ Unlike an ordinary
space, a spectral presheaf has no ``points'' or global elements
\cite{key-3}, in concordance with the Kochen-Specker theorem. In
this way, the pure states $\omega$ (on $\widehat{\mathcal{A}}_{sp.}\doteq R_{QG}\left[\widehat{\mathcal{A}}_{SP.}\right]$,
not on $\widehat{\mathcal{A}}_{SP.}$), the space's ``points'',
are mapped to clopen sub-objects of $\underline{\varSigma}_{R_{QG}}$,
i.e. our ``quantum points'' will be much like regions, in that they
will have nonempty interiors. This will go hand in hand with the results
in the next sections, which deal with the metric aspect of this. Furthermore,
there's an obvious quantum nature in the points, since now we can
have incompatible points (that is, points that correspond to non-commuting
projectors in the Hilbert space; this is only possible because the
space algebra is non-commutative)}$.\,\blacksquare$

\selectlanguage{spanish}%
$\;$

\selectlanguage{english}%
$\mathbf{Definition\;2.6}$:\foreignlanguage{spanish}{ The analogue
of a given classical metric $h_{ab}$ will be given by a $3-$(spectral)
dimensional \emph{real first order spectral triple} 
\[
\mathcal{T}_{D_{QG}}^{R_{QG}}\doteq\left(R_{QG}\left[\widehat{\mathcal{A}}_{SP.}\right],\mathcal{H}_{R_{QG}},D_{QG}\right).\,\blacksquare
\]
(Of course, the triple does contain differential geometric information,
but it's encoded algebraically and functional analytically rather
than in a space picture.)}

$\;$

$\mathbf{Remark\;2.2}$:\foreignlanguage{spanish}{ In particular,
one could try to classify all the \emph{Dirac-like operators} $D_{QG}$
there compatible with the algebra representation and \emph{inner product}
$\left(\cdot,\cdot\right)_{\mathcal{H}_{R_{QG}}}$ of its Hilbert
space; \emph{each of these operators would be an analogue of a given}
$h_{ab}$ (since $\widehat{\mathcal{A}}_{sp.}\doteq R_{QG}\left[\widehat{\mathcal{A}}_{SP.}\right]$),
and they could be used to calculate \emph{non-commutative} space intervals
(i.e. distances) and \emph{non-commutative} volume integrals, which,
given the interpretations we made, \emph{are the genuinely quantum
distances and volumes}, since they are based on the \emph{quantized
space algebra$.\,\blacksquare$}}

$\;$

$\mathbf{Definition\;2.7}$:\foreignlanguage{spanish}{\emph{ The complete
quantum analogue of the classical space picture $(\varSigma,h_{ab})$
will be the pair} 
\[
(\underline{\varSigma}_{R_{QG}},\mathcal{T}_{D_{QG}}^{R_{QG}}).\,\blacksquare
\]
}

\selectlanguage{spanish}%
$\;$

We call all the ideas in sections 1 and 2.1 ``\emph{NCR space(-time})''
(i.e. Non-Commutative Relational space(-time)). These ideas are very
general and \emph{independent} of their concrete implementation, that
is, of the selection of \foreignlanguage{english}{a particular, concrete,
background independent Poisson sub-algebra $\mathcal{A}_{GR}$ of
phase space functionals} as required in Definition 2.1. What follows
next are proposals for their \emph{concrete} implementation. This
step is quite non-trivial and currently we only have a partial answer
in the sense that we're only able to build the $3-$dimensional spatial
part of spacetime without the temporal one (this is a consequence
of the fact that the formalism of LQG is based on a $3+1$ Hamiltonian
formalism instead of the so-called ``covariant'' one; but the $3+1$
formalism is, to date, the only one of these which has been quantized
in a rigorous mathematical way, and this is one of the great results
of LQG).

\selectlanguage{english}%
$\;$

\subsubsection*{2.2. AQG Relational Space}

$\;$

\selectlanguage{spanish}%
First, we need the quantized phase space algebra $\widehat{\mathcal{A}}_{QG}$.
The immediate choice would be the kinematic quantum algebra $\widehat{\mathcal{A}}_{LQG}$
from LQG. Unfortunately, this algebra makes a heavy use of a continuum
background manifold $\Sigma$; for example, the ``electric-flux''
elements $\widehat{\mathrm{E}}(S,\boldsymbol{k})\in\widehat{\mathcal{A}}_{LQG}$,
where $\boldsymbol{k}\in C^{\infty}(S,\mathfrak{su}(2))$ and $S\subset\varSigma$
is a \emph{surface}, are \emph{parameterized} by the surfaces $S$,
which are submanifolds of $\varSigma$. Thus, if we use that algebra,
we would be falling into the physical contradictions which we were
seeking to avoid with all this approach.

$\;$

Now, the principle of canonical quantization suggests that \emph{all}\footnote{\selectlanguage{english}%
It's a common myth that properties such as the quantum spin are ``purely''
quantum and don't have a classical analogue: this has proven to be
false by results from Geometric Quantization and other more modern
approaches \cite{key-10}.\selectlanguage{spanish}%
} the properties of the quantum theory descend to the classical one,
since one can reconstruct the quantum theory only by knowing the structure
of the phase space of the classical theory. On the other hand, one
suspects that the continuum manifold only appears in the classical
limit, and, therefore, any properties in the classical theory that
depend on it \emph{cannot} be considered genuine properties that descended
from the fundamental quantum theory, but rather, \emph{spurious} properties
introduced by the classical limit and which only exist at that level.
Thus, we adopt the following \emph{hypothesis}:

$\;$

\selectlanguage{english}%
$\mathbf{Hypothesis\;2.2}$:\foreignlanguage{spanish}{\emph{ If one
takes the set of classical properties in classical phase space that
allows to relationally reconstruct the classical manifold, and strips
out all what makes reference to the continuum manifold, what remains
are the true fundamental properties that will allow to relationally
reconstruct the quantum space in the quantum theory once these properties
have been canonically quantized$.\,\blacksquare$}}

\selectlanguage{spanish}%
$\;$

(This should clarify the process ``Dis.'' of Definition 2.2, which
stands for a kind of \textit{fundamental} discretization, i.e. one
without a cutoff parameter, since what's being taken out is considered
simply to be unnecessary to begin with, rather than something we just
take out for the sake of approximation.)

$\;$

A way to implement this is perhaps provided by a close cousin of LQG
called Algebraic Quantum Gravity (AQG) \cite{key-8}, to which we
now turn our attention.

\selectlanguage{english}%
$\;$

$\mathbf{Definition\;2.8}$:\foreignlanguage{spanish}{\emph{ }\cite{key-2}
In LQG, phase space is coordinatized by the canonical pairs $\left[\boldsymbol{A}_{a},\boldsymbol{E}^{a}\right]$,
}where $\boldsymbol{A}_{a}$ is a $SU(2)$ Yang-Mills connection on
$\varSigma$ and $\boldsymbol{E}^{a}$ its associated ``Yang-Mills
electric field'' (both fields are $\mathfrak{su}(2)-$valued, as
represented by the bold font). Then, in LQG,\foreignlanguage{spanish}{
the classical $\mathcal{A}_{LQG}$ is generated by the following phase
space properties}\footnote{\selectlanguage{spanish}%
Where $*E_{ab}^{\left(i\right)}\doteq e_{abc}E_{\left(i\right)}^{c},\,i=1,2,3.$,
\foreignlanguage{english}{$e$ is a path, $S\subset\varSigma$ a surface,
and $\boldsymbol{k}\in C^{\infty}(S,\mathfrak{su}(2))$.}\selectlanguage{english}%
}\foreignlanguage{spanish}{:}

\selectlanguage{spanish}%
\[
\boldsymbol{\mathrm{H}}(e)\left(\left[\boldsymbol{A}_{a},\boldsymbol{E}^{a}\right]\right)\doteq\mathcal{P}\,\mathrm{exp}\int_{e}\boldsymbol{A}_{a}\quad\mathrm{and}\quad\mathrm{E}(S,\boldsymbol{k})\left(\left[\boldsymbol{A}_{a},\boldsymbol{E}^{a}\right]\right)\doteq\int_{S}*\boldsymbol{E}_{ab}\cdot\boldsymbol{k}.\,\blacksquare
\]

$\;$

\selectlanguage{english}%
$\mathbf{Proposition\;2.1}$: The variables in Definition 2.8 \emph{can}
be used to obtain $C^{\infty}(S)$ via $R_{h}$.

$\;$

\emph{Proof}: pick any $\boldsymbol{k}_{0}\in C^{\infty}(S,\mathfrak{su}(2))$
such that $\boldsymbol{k}_{0}\neq0$ in \emph{all} of $S$ and write

\[
\boldsymbol{k}\doteq f\boldsymbol{k}_{0},\,f\in C^{\infty}(S).
\]
Since $*\boldsymbol{E}_{ab}\cdot\boldsymbol{k}$ is a $2-$form on
$S$, we have

\[
*\boldsymbol{E}_{ab}\cdot\boldsymbol{k}=f*\boldsymbol{E}_{ab}\cdot\boldsymbol{k}_{0}
\]

\[
=f\,f_{\boldsymbol{E}^{a},\boldsymbol{k}_{0}}\boldsymbol{\epsilon}_{ab}(q),
\]
where $f_{\boldsymbol{E}^{a},\boldsymbol{k}_{0}}\in C^{\infty}(S)$
depends on $\boldsymbol{E}^{a}$ and $\boldsymbol{k}_{0}$ (and $q$
comes from the $h$ in $\left[h_{ab},\pi^{ab}\right]$ determined
by the point $\left[\boldsymbol{A}_{a},\boldsymbol{E}^{a}\right]$;
see \cite{key-2} for how these variables transform into each other).
For $f_{1},f_{2}\in C^{\infty}(S)$, consider the product $._{S}$
defined by:

\[
\left(\mathrm{E}(S,f_{1}\boldsymbol{k}_{0})\cdot_{S}\mathrm{E}(S,f_{2}\boldsymbol{k}_{0})\right)\left(\left[\boldsymbol{A}_{a},\boldsymbol{E}^{a}\right]\right)\doteq\mathrm{E}(S,f_{1}f_{2}\boldsymbol{k}_{0}),\,\forall\left[\boldsymbol{A}_{a},\boldsymbol{E}^{a}\right]\in X
\]

\[
=\int_{S}f_{1}f_{2}\,f_{\boldsymbol{E}^{a},\boldsymbol{k}_{0}}\boldsymbol{\epsilon}(q).
\]
With this product and for \emph{each} $\boldsymbol{k}_{0}$, define
the algebras:

\[
\mathcal{W}_{\boldsymbol{k}_{0}}\doteq\left(\left\{ \mathrm{E}(S,f\boldsymbol{k}_{0})\right\} _{f\in C^{\infty}(S)},\cdot_{S}\right).
\]
Now, clearly, since $\boldsymbol{k}_{0}\neq0$ in \emph{all} of $S$,
the algebras for different $\boldsymbol{k}_{0}$ are \emph{all isomorphic
and equivalent to} $C^{\infty}(S)$ (since $f\boldsymbol{k}_{0}=f'\boldsymbol{k}_{0}$
implies $f=f'$). In this way, for any solution characterized by $\left[\boldsymbol{A}_{a},\boldsymbol{E}^{a}\right]$
such that $\boldsymbol{E}^{a}\neq0$ in \emph{all} of $S$, there's
at least one $\boldsymbol{k}_{0}$ such that\footnote{Indeed, just pick $\boldsymbol{E}^{a}$ (normalized) as one of the
elements of a triad basis of both the tangent spaces and $\mathfrak{su}(2)$,
and then, in that basis, $\left(k_{0}\right)_{(\vartriangle)}=\frac{\sqrt{q}}{E_{(\vartriangle)}^{*}}$,
since $E_{(\vartriangle)}^{*}\neq0$ in $S$.}

\[
f_{\boldsymbol{E}^{a},\boldsymbol{k}_{0}}=1\,\mathrm{in\,all\,of}\,S,
\]
and, in this way, the algebra for that $\boldsymbol{k}_{0}$ gets
\emph{bijectively} mapped to $C^{\infty}(S)$ by the relational representation
$R_{h}$. Thus, for any point $\left[\boldsymbol{A}_{a},\boldsymbol{E}^{a}\right]$
such that $\boldsymbol{E}^{a}\neq0$ in \emph{all} of $S$, there's
always one of those algebras that gets bijectively mapped to $C^{\infty}(S)$
by the relational representation $R_{h}$; but, since all the algebras
are isomorphic, this means that the relational representation induces
the algebra $C^{\infty}(S)$ of the surface $S$ from a \emph{single}
algebra of phase space functions for \emph{any} of those points $\left[\boldsymbol{A}_{a},\boldsymbol{E}^{a}\right]$$.\,\square$

\selectlanguage{spanish}%
$\;$

\selectlanguage{english}%
$\mathbf{Definition\;2.9}$:\foreignlanguage{spanish}{ \cite{key-8}
In AQG, one considers an \emph{abstract algebraic (countably infinite)
graph $\alpha$} (of distinguishable edges $\mathrm{e}$) and an \emph{embedding}
$\gamma\doteq\psi(\alpha)$ on $\varSigma$ that makes it \emph{dual}
to a \emph{triangulation} $\gamma^{*}$. Thus, for each $e\doteq\psi(\mathrm{e})$
there's a \emph{unique face} $S_{e}$ in $\gamma^{*}$ which intersects
$e$, and its does so only at an interior point $p_{e}$ of both $S_{e}$
and $e$. For each $x\in S_{e}$, choose a path $\rho_{e}(x)$ which
starts at $\partial e$, runs along $e$ until $p_{e}$, and then
whitin $S_{e}$ until $x$. Then, we define the following phase space
functions:}

\selectlanguage{spanish}%
\[
\boldsymbol{\mathrm{H}}(\mathrm{e})\left(\left[\boldsymbol{A}_{a},\boldsymbol{E}^{a}\right]\right)\doteq\mathcal{P}\,\mathrm{exp}\int_{\psi(\mathrm{e})}\boldsymbol{A}_{a}
\]
and

\selectlanguage{english}%
\[
\mathrm{E}_{(j)}(\mathrm{e})\left(\left[\boldsymbol{A}_{a},\boldsymbol{E}^{a}\right]\right)\doteq\int_{S_{\psi(\mathrm{e})}}e_{abc}(x)\left[\boldsymbol{H}_{\rho_{\psi(\mathrm{e})}(x)}(\boldsymbol{A}_{a})\boldsymbol{E}^{c}(x)\boldsymbol{H}_{\rho_{\psi(\mathrm{e})}(x)}^{-1}(\boldsymbol{A}_{a})\right]\cdot\tau_{j},
\]
\foreignlanguage{spanish}{where $\boldsymbol{H}_{\rho_{e}(x)}(\boldsymbol{A}_{a})$
is the holonomy and $\left\{ \tau_{j}\right\} _{j=1,2,3.}$ is a basis
of $\mathfrak{su}(2)$ (of course, $\tau_{i}\cdot\tau_{j}=\delta_{ij}$
and $\mathrm{tr}\,(\tau_{i}\tau_{j})=2\delta_{ij}$). }

\selectlanguage{spanish}%
We denote by $\mathcal{A}_{AQG}$ the algebra generated (with \emph{real}
coefficients) by the above phase space functions$.\,\blacksquare$

\selectlanguage{english}%
$\;$

$\mathbf{Lemma\;2.1}$:\foreignlanguage{spanish}{ \cite{key-8} The
Poisson brackets for the variables in Definition 2.9 are given by:}

\selectlanguage{spanish}%
\[
\left\{ \boldsymbol{\mathrm{H}}(\mathrm{e}),\boldsymbol{\mathrm{H}}(\mathrm{e}')\right\} =0,
\]

\[
\left\{ \mathrm{E}_{(j)}(\mathrm{e}),\boldsymbol{\mathrm{H}}(\mathrm{e}')\right\} =-\delta_{\mathrm{ee}'}\tau_{j}\boldsymbol{\mathrm{H}}(\mathrm{e}),
\]

\[
\left\{ \mathrm{E}_{(j)}(\mathrm{e}),\mathrm{E}_{(k)}(\mathrm{e}')\right\} =\delta_{\mathrm{ee}'}\sum_{l=1}^{3}\varepsilon_{jkl}\mathrm{E}_{(l)}(\mathrm{e}).\,\square
\]

$\;$

\selectlanguage{english}%
$\mathbf{Lemma\;2.2}$: ($G=SU(2)$, in our case.) The Poisson bracket
relations of Lemma 2.1 for $\mathrm{e}=\mathrm{e}'$ are equivalent
to those of the phase space defined by the cotangent bundle $T^{*}G\cong\mathfrak{g}^{*}\times G$
with the Poisson brackets given by the so-called semidirect product
Poisson structure (see \cite{key-10} for some details), where the
$\mathrm{E}_{(j)}(\mathrm{e})$ are simply the generators of $\mathfrak{su}(2)$\foreignlanguage{spanish}{$.\,\square$}

$\;$

$\mathbf{Remark\;2.3}$:\foreignlanguage{spanish}{ As one can see,
\emph{both the variables and their Poisson brackets only depend on
the abstract algebraic graph and all reference to the continuum manifold
$\varSigma$ has completely vanished}$.\,\blacksquare$}

\selectlanguage{spanish}%
$\;$

\selectlanguage{english}%
$\mathbf{Lemma\;2.3}$:\foreignlanguage{spanish}{ The variables of
the proof of Propostion 2.1, $\mathrm{E}(S,f\boldsymbol{k}_{0})$
(considering the set of \emph{all} the elements from \emph{all} the
algebras $\mathcal{W}_{\boldsymbol{k}_{0}}\doteq\left(\left\{ \mathrm{E}(S,f\boldsymbol{k}_{0})\right\} _{f\in C^{\infty}(S)},\cdot_{S}\right)$
that were defined), subjected to the process of }Definition 2.2/Hypothesis
2.2\foreignlanguage{spanish}{, result in the variables of Definition
2.9.}

$\;$

\emph{Proof}:\foreignlanguage{spanish}{ the process in }Definition
2.2/Hypothesis 2.2\foreignlanguage{spanish}{ amounts to only retaining
the generic constant functions $c$ instead of the general $f$ and
a constant generic $\mathfrak{su}(2)$ matrix $\boldsymbol{C}=\sum_{j=1}^{3}C_{j}\tau_{j}$
instead of the general function $\boldsymbol{k}_{0}$. Thus,}

\selectlanguage{spanish}%
\[
\mathrm{E}(S_{e},c\boldsymbol{C})\left(\left[\boldsymbol{A}_{a},\boldsymbol{E}^{a}\right]\right)\doteq\sum_{j=1}^{3}cC_{j}\int_{S_{e}}*E_{ab}^{\left(j\right)}
\]

\[
=\sum_{j=1}^{3}cC_{j}\mathrm{E}(S_{e},\tau_{j})\left(\left[\boldsymbol{A}_{a},\boldsymbol{E}^{a}\right]\right),
\]
which indicates that the functions to quantize are of the form (absorbing
$c$ into $\boldsymbol{C}$)

\[
\mathrm{E}(S_{e},\boldsymbol{C})=\sum_{j=1}^{3}C_{j}\mathrm{E}_{(j)}(\mathrm{e}),
\]
which gives the \emph{linearly generated} $\mathrm{E}$ part of the
previous algebra $\mathcal{A}_{AQG}$ .$\,\square$

$\;$

Now, in AQG, the abstract quantized commutation relations of $\mathcal{\widehat{A}}_{AQG}$
are:

\[
\left[\widehat{\boldsymbol{\mathrm{H}}}(\mathrm{e}),\widehat{\boldsymbol{\mathrm{H}}}(\mathrm{e}')\right]=0,
\]

\[
\left[\widehat{\mathrm{E}}_{(j)}(\mathrm{e}),\widehat{\boldsymbol{\mathrm{H}}}(\mathrm{e}')\right]=-\mathrm{i}\delta_{\mathrm{ee}'}\tau_{j}\widehat{\boldsymbol{\mathrm{H}}}(\mathrm{e}),
\]

\[
\left[\widehat{\mathrm{E}}_{(j)}(\mathrm{e}),\widehat{\mathrm{E}}_{(k)}(\mathrm{e}')\right]=\mathrm{i}\delta_{\mathrm{ee}'}\sum_{l=1}^{3}\varepsilon_{jkl}\widehat{\mathrm{E}}_{(l)}(\mathrm{e}).
\]

$\;$

\selectlanguage{english}%
$\mathbf{Definition\;2.10}$:\foreignlanguage{spanish}{ We define
$\mathcal{F}_{Dis}\equiv\mathcal{A}_{SP.}\equiv\mathcal{A}_{\mathrm{e}}\subset\mathcal{A}_{AQG}$
as the set linearly generated by the $\mathrm{E}_{(j)}(\mathrm{e})$.
Now, we use the $\widehat{\mathrm{E}}_{(j)}(\mathrm{e})$ variables
from AQG to define, following Definition 2.2, the quantum algebra
\[
\widehat{\mathcal{A}}_{SP.}\equiv\widehat{\mathcal{A}}_{\widehat{S_{e}}}\doteq\widehat{\mathfrak{su}(2)}\subset U(\mathfrak{su}(2))
\]
(where the generators of $\widehat{\mathfrak{su}(2)}$ in $U(\mathfrak{su}(2))$
are now $-\mathrm{i}\widehat{\mathrm{E}}_{(j)}(\mathrm{e})$) which
corresponds to the quantization $\widehat{S_{e}}$ of the $1-$punctured
surface $S_{e}$$.\,\blacksquare$ }

\selectlanguage{spanish}%
$\;$

\selectlanguage{english}%
$\mathbf{Lemma\;2.4}$:\foreignlanguage{spanish}{ \cite{key-8} A
\emph{concrete self-adjoint representation} $R_{AQG}$ of $\mathcal{\widehat{A}}_{AQG}$
is implemented on the \emph{infinite tensor product Hilbert space}}

\selectlanguage{spanish}%
\[
\mathcal{H}^{\otimes_{\infty}}\doteq\otimes_{\mathrm{e}}\mathcal{H}^{\mathrm{e}},\quad\mathrm{where}\quad\mathcal{H}^{\mathrm{e}}\doteq L^{2}(SU(2),\nu)
\]
(which is the closure of the finite linear span of vectors of the
form $\otimes_{f}\doteq\otimes_{\mathrm{e}}f_{\mathrm{e}}$, where
$f_{\mathrm{e}}\in\mathcal{H}^{\mathrm{e}}$; $\nu$ is the Haar measure
of $SU(2)$) by the operators

\[
\widehat{\boldsymbol{\mathrm{H}}}(\mathrm{e})\otimes_{f}\doteq\left[\widehat{\boldsymbol{\mathrm{H}}}(\mathrm{e})f_{\mathrm{e}}\right]\otimes\left[\otimes_{\mathrm{e}'\neq\mathrm{e}}f_{\mathrm{e}'}\right],
\]

\[
\widehat{\mathrm{E}}_{(j)}(\mathrm{e})\otimes_{f}\doteq\left[\widehat{\mathrm{E}}_{(j)}(\mathrm{e})f_{\mathrm{e}}\right]\otimes\left[\otimes_{\mathrm{e}'\neq\mathrm{e}}f_{\mathrm{e}'}\right],
\]
where

\[
\left[\widehat{\boldsymbol{\mathrm{H}}}(\mathrm{e})f_{\mathrm{e}}\right](g)\doteq gf_{\mathrm{e}}(g),
\]

\[
\left[\widehat{\mathrm{E}}_{(j)}(\mathrm{e})f_{\mathrm{e}}\right](g)\doteq\mathrm{i}\left[\frac{\mathrm{d}}{\mathrm{d}t}f_{\mathrm{e}}(ge^{-t\tau_{j}})\right]\mid_{t=0}.\,\square
\]

$\;$

Since we are studying the quantization $\widehat{S_{e}}$ of the punctured
surface $S_{e}$, we now turn our attention to the algebra $\widehat{\mathcal{A}}_{\widehat{S_{e}}}$
for the $\mathrm{e}$ part and its representation space $\mathcal{H}^{\mathrm{e}}\doteq L^{2}(SU(2),\nu)$
(we will omit the $\mathrm{e}$ from now on).

$\;$

\selectlanguage{english}%
Before continuing, we're going to need the following examples:

$\;$

$\mathbf{Example\;2.1.1}$: (cf. \cite{key-13} for details and proofs.)
Consider a (compact) Riemannian symmetric space $(M,\nu)$ ($\mathrm{dim}\,M=m$)
with isotropy (Lie) group $G$. As is well known, if $K$ is the isotropy
(or stabilizer) group of a fixed point $p\in M$, then $M$ can be
identified with the homogeneous coset space, that is, $M\cong\frac{G}{K}$.
The Lie algebra $\mathfrak{g}$ of $G$ can then be split into $\mathfrak{g}=\mathfrak{k}+\mathfrak{m}$
(where $[\mathfrak{k},\mathfrak{k}]\subset\mathfrak{k}$, $[\mathfrak{k},\mathfrak{m}]\subset\mathfrak{m}$,
and $[\mathfrak{m},\mathfrak{m}]\subset\mathfrak{k}$). Now, to fix
a homogeneous spin structure on $M$ is to have a homomorphism $\widetilde{\mathrm{Ad}}:K\,\longrightarrow Spin\,(\mathfrak{m})$
such that the following diagram commutes (where $\lambda$ is the
usual covering map):

\[
\xymatrix{ & Spin\,(\mathfrak{m})\ar[d]_{\lambda}\\
K\ar[ur]^{\widetilde{\mathrm{Ad}}}\ar[r]^{\mathrm{Ad}} & SO(\mathfrak{m})
}
\]
Let $\kappa:Spin\,(\mathfrak{m})\,\longrightarrow GL(\triangle)$
be the spin representation (where $\triangle$ is the vector space
of spinors). A spinor field $\check{\psi}$ on $M$ is identified
with a function $\check{\psi}:G\,\longrightarrow\triangle$ that satisfies
the invariance condition $\check{\psi}(gk)=\kappa\widetilde{\mathrm{Ad}}(k^{-1})\check{\psi}(g)$.
For $X\in\mathfrak{g}$, the left invariant vector field $X$ on $G$
is defined in the standard way, $X(\check{\psi})(g)\doteq\frac{\mathrm{d}}{\mathrm{d}t}\left[\check{\psi}(ge^{-tX})\right]\mid_{t=0}$.
With this, we can (locally) define the Dirac operator as 
\[
\check{\cancel{D}}\check{\psi}\doteq-\mathrm{i}\sum_{j=1}^{m}\gamma^{j}X_{j}(\check{\psi}).
\]
Then, if $\Omega_{G}=-\sum_{j=1}^{dim\,G}X_{j}^{2}$ is the Casimir
operator of $G$, the following important result holds: $\check{\cancel{D}^{2}}=\Omega_{G}+\frac{1}{8}\mathsf{s}.$
The power of this formula lies in that it allows us to calculate the
eigenvalues of $\check{\cancel{D}^{2}}$ and $\mid\check{\cancel{D}}\mid$
purely by means of the representation theory of $G$. Furthermore,
the canonical, \emph{commutative} spectral triple of Example 1.1,
$\left(C^{\infty}(M),L^{2}(\mathcal{S},\boldsymbol{\epsilon}(g_{ab})),\cancel{D}_{g}\right)$,
can now be written as $\left(C^{\infty}(\frac{G}{K}),L^{2}(\check{\mathcal{S}},\nu),\check{\cancel{D}}\right)$$.\,\blacksquare$

$\;$

$\mathbf{Example\;2.1.2}$: In the particular case of $G=SU(2)$,
we have two examples of interest: 
\begin{enumerate}
\item The first one is $K=\{e\}$, which means that $M\cong\frac{G}{K}=G=SU(2)\cong\mathbb{S}^{3}$,
i.e. the $3-$sphere. The Dirac operator is 
\[
\check{\cancel{D}}_{\mathbb{S}^{3}}\check{\psi}\doteq-\mathrm{i}\sum_{j=1}^{3}\sigma^{j}X_{j}(\check{\psi}),
\]
where $\sigma^{j}$ are the standard Pauli matrices\footnote{We use the convention: $\sigma^{1}\doteq\left(\begin{array}{cc}
0 & 1\\
1 & 0
\end{array}\right)$, $\sigma^{2}\doteq\left(\begin{array}{cc}
0 & -\mathrm{i}\\
\mathrm{i} & 0
\end{array}\right)$, $\sigma^{3}\doteq\left(\begin{array}{cc}
1 & 0\\
0 & -1
\end{array}\right)$.}. Note that, in this case, since $\triangle=\mathbb{C}^{2^{1}}$ ($3=n=2m+1$),
we have $L^{2}(\check{\mathcal{S}}_{\mathbb{S}^{3}},\nu)\cong\mathbb{C}^{2}\otimes L^{2}(SU(2),\nu)$. 
\item The second example is $K=U(1)$, which means $M\cong\frac{SU(2)}{U(1)}\cong\mathbb{S}^{2}$,
i.e. the $2-$sphere. The Dirac operator is 
\[
\check{\cancel{D}}_{\mathbb{S}^{2}}\check{\psi}\doteq-\mathrm{i}\sum_{j=1}^{2}\sigma^{j}X_{j}(\check{\psi}).
\]
In this case, $L^{2}(\check{\mathcal{S}}_{\mathbb{S}^{2}},\nu)\cong\mathbb{C}^{2^{1}}\otimes L^{2}(\mathbb{S}^{2},\nu)$
($2=n=2m$). 
\end{enumerate}
$\;$

$\mathbf{Lemma\;2.5}$: (cf. \cite{key-14} for some details.) Up
to normalization constants, the spinors\footnote{\selectlanguage{spanish}%
By the Peter-Weyl theorem \cite{key-12}, we know that the functions\foreignlanguage{english}{
$SU(2)\ni g\,\longmapsto\varPi^{\left(s\right)}\left(g\right)_{ij},\,i,j=1,...,\left(2s+1\right),\,\forall s\in\frac{1}{2}\mathbb{N}_{0},$}
given by the matrix elements of the irreducible representations of
$SU(2)$ on the (necessarily, \emph{finite} dimensional) Hilbert spaces
$\mathcal{H}_{(s)}$ (where $dim\,\mathcal{H}_{(s)}=2s+1$), comprise
a basis of the \emph{separable} $L^{2}(SU(2),\nu)$. Also, the same
theorem states that $L^{2}(SU(2),\nu)$ decomposes into a direct sum
of spaces on which the left regular representation $\left[\mathfrak{L}_{h}\left(f_{\mathrm{e}}\right)\right](g)\doteq f_{\mathrm{e}}(gh^{-1}),\,\forall h,g\in SU(2),\,\forall f_{\mathrm{e}}\in L^{2}(SU(2),\nu)$
of $SU(2)$ on $L^{2}(SU(2),\nu)$ {[}note that \emph{this} representation
\emph{is the one used to define} $\widehat{\mathrm{E}}_{(j)}(\mathrm{e})$
above{]} is \emph{irreducible}. \emph{All} irreducible representations
of $SU(2)$ appear in this decomposition and the one for spin $s$
(which appears $\left(2s+1\right)$ times in the decomposition) is
realized by the matrices $\varPi^{\left(s\right)}\left(g\right)$
acting on the sub-space of $L^{2}(SU(2),\nu)$ spanned by the previous
$\varPi_{ij}^{\left(s\right)},\,i,j=1,...,\left(2s+1\right)$ (or,
in a different convention, $\varPi_{m\mu}^{\left(s\right)}$, $m,\mu=-s,-s+1,...,s-1,s$),
and isomorphic to $\mathcal{H}_{(s)}$, i.e. \foreignlanguage{english}{ 
\[
L^{2}(SU(2),\nu)\cong\mathbb{C}\oplus\mathcal{H}_{(\frac{1}{2})}\oplus\mathcal{H}_{(\frac{1}{2})}\oplus...\oplus\underset{\left(2s+1\right)}{\underbrace{\mathcal{H}_{(s)}\oplus...\oplus\mathcal{H}_{(s)}}}\oplus...
\]
(}the $j_{0}-\mathrm{th,}\,1\leq j_{0}\leq\left(2s+1\right),$ space
$\mathcal{H}_{(s)}$ in $\underset{\left(2s+1\right)}{\underbrace{\mathcal{H}_{(s)}\oplus...\oplus\mathcal{H}_{(s)}}}$
is spanned by $\left\{ \varPi_{ij_{0}}^{\left(s\right)}\right\} _{1\leq i\leq\left(2s+1\right)}$).\selectlanguage{english}%
} 
\[
Y_{s'\mu'}^{+}\doteq\left(\begin{array}{c}
\varPi_{-\frac{1}{2},\,\mu'}^{\left(s'\right)}\mid_{(-\varphi,\theta,\varphi)}\\
\mathrm{i}\varPi_{\frac{1}{2},\,\mu'}^{\left(s'\right)}\mid_{(-\varphi,\theta,\varphi)}
\end{array}\right),\qquad Y_{s'\mu'}^{-}\doteq\left(\begin{array}{c}
-\varPi_{-\frac{1}{2},\,\mu'}^{\left(s'\right)}\mid_{(-\varphi,\theta,\varphi)}\\
\mathrm{i}\varPi_{\frac{1}{2},\,\mu'}^{\left(s'\right)}\mid_{(-\varphi,\theta,\varphi)}
\end{array}\right),
\]
$\forall s'\in\mathbb{N}_{0}+\frac{1}{2}$, $\forall\mu'\in\underset{\left(2s'+1\right)}{\underbrace{\{-s',-s'+1,...,s'-1,s'\}}}$,
where the $\varPi_{\pm\frac{1}{2},\,\mu'}^{\left(s'\right)}\mid_{(-\varphi,\theta,\varphi)}e^{\mp i(\psi+\varphi)/2}=\varPi_{\pm\frac{1}{2},\,\mu'}^{\left(s'\right)}\mid_{(\psi,\theta,\varphi)}$
form an \emph{eigenbasis of} $\check{\cancel{D}}_{\mathbb{S}^{2}}$
\emph{in} $L^{2}(\check{\mathcal{S}}_{\mathbb{S}^{2}},\nu)$, such
that

\[
\check{\cancel{D}}_{\mathbb{S}^{2}}Y_{s'\mu'}^{+}=\left(s'+\frac{1}{2}\right)Y_{s'\mu'}^{+},\qquad\check{\cancel{D}}_{\mathbb{S}^{2}}Y_{s'\mu'}^{-}=-\left(s'+\frac{1}{2}\right)Y_{s'\mu'}^{-}.\,\square
\]
In this way, the eigenvalues of $\check{\cancel{D}_{\mathbb{S}^{2}}^{2}}$
are such that $\left(s'+\frac{1}{2}\right)^{2}=s'(s'+1)+\frac{1}{4}$,
so we can then recognize $s'(s'+1)$ as the eigenvalues of the Casimir
element $\Omega_{G}$ and the value for the scalar curvature $\mathsf{s}_{\,_{\mathbb{S}^{2}}}=2$
of $(\mathbb{S}^{2},\nu)$. Noting that $\mathsf{s}_{\,_{\mathbb{S}^{2}}}=\frac{2}{\rho_{\mathbb{S}^{2}}^{2}}$,
where $\rho_{\,_{\mathbb{S}^{2}}}$ is the radius of $(\mathbb{S}^{2},\nu)$,
we will write the eigenvalues $d_{s'}$ of $\check{\cancel{D}}_{\mathbb{S}^{2}}$
as 
\[
d_{s'}=\sqrt{s'(s'+1)+\left(\frac{1}{4\rho_{\,_{\mathbb{S}^{2}}}^{2}}\right)}.\,\blacksquare
\]

$\;$

Thus, these examples show that we have at least two Dirac operators
that we can define on the representation space of AQG. But the one
we will use is inspired in $\check{\cancel{D}}_{\mathbb{S}^{2}}$,
i.e. that one corresponding to the $2-$sphere, since we are quantizing
a surface.

\selectlanguage{spanish}%
$\;$

\selectlanguage{english}%
$\mathbf{Definition\;2.11}$:\foreignlanguage{spanish}{ For each $s\in\frac{1}{2}\mathbb{N}_{0}$,
we define $\mathcal{H}_{\mu}^{(s)}\subset\underset{\left(2s+1\right)}{\underbrace{\mathcal{H}_{(s)}\oplus...\oplus\mathcal{H}_{(s)}}}$
to be the space spanned by $\left\{ \varPi_{m\mu}^{\left(s\right)}\right\} _{1\leq m\leq\left(2s+1\right)}$.
Then, we reduce to $\mathcal{H}_{\mu}^{(s)}$ the action of the algebra
representation $R_{AQG}\left[\widehat{\mathfrak{su}(2)}\right]$.
We denote the result as $R_{AQG}\left[\widehat{\mathfrak{su}(2)}\right]_{\mu}^{(s)}$$.\,\blacksquare$}

$\;$

$\mathbf{Definition\;2.12}$: We define 
\[
D_{\mu}^{(s)}\doteq-\mathrm{i}\sum_{j=1}^{2}\sigma^{j}X_{j}
\]
on $\check{\mathcal{H}}_{\mu}^{(s)}\doteq\mathbb{C}^{2}\otimes\mathcal{H}_{\mu}^{(s)}$,
$\forall s\in\frac{1}{2}\mathbb{N}_{0}$, where $-\mathrm{i}\widehat{\mathrm{E}}_{(j)}(\mathrm{e})=X_{j}$
(which are therefore \emph{skew-adjoint} rather than self-adjoint,
like the $\widehat{\mathrm{E}}_{(j)}$)$.\,\blacksquare$

$\;$

$\mathbf{Proposition\;2.2}$: On the spaces $\check{\mathcal{H}}_{\mu}^{(s)}$,
we have 
\[
\left(D_{\mu}^{(s)}\right)^{2}u_{m\mu}^{\pm}=d_{s,m}^{2}u_{m\mu}^{\pm},
\]

\[
u_{m\mu}^{\pm}=\left(\begin{array}{c}
\pm\varPi_{-m\mu}^{\left(s\right)}\\
\mathrm{i}\varPi_{m\mu}^{\left(s\right)}
\end{array}\right),\;d_{s,m}^{2}=s(s+1)-m^{2}-m,\;\forall s\in\frac{1}{2}\mathbb{N}_{0},\;\forall m\in\underset{\left(2s+1\right)}{\underbrace{\{-s,-s+1,...,s-1,s\}}}.
\]

$\;$

\emph{Proof}: we have:

\[
D^{2}=-(\sigma^{1}X_{1}+\sigma^{2}X_{2})^{2}
\]

\[
=-I_{2\times2}X_{1}^{2}-I_{2\times2}X_{2}^{2}-\sigma^{1}\sigma^{2}X_{1}X_{2}-\sigma^{2}\sigma^{1}X_{2}X_{1}
\]

\[
=-I_{2\times2}X_{1}^{2}-I_{2\times2}X_{2}^{2}-\sigma^{1}\sigma^{2}[X_{1},X_{2}]
\]

\[
=I_{2\times2}\Omega_{G}-I_{2\times2}(-X_{3}^{2})-\sigma^{3}(\mathrm{i}X_{3});
\]
since $(\mathrm{i}X_{3})\varPi_{m\mu}^{\left(s\right)}=-m\varPi_{m\mu}^{\left(s\right)}$,
$\forall\mu$, the $d_{s,m}^{2}$ are eigenvalues of $D^{2}$ for
$u_{m\mu}^{\pm}$$.\,\square$

$\;$

$\mathbf{Remark\;2.4}$: \foreignlanguage{spanish}{In this way}, by
using the relation $\varPi_{\pm\frac{1}{2},\mu}^{\left(s'\right)}\mid_{(-\varphi,\theta,\varphi)}\,\longmapsto\varPi_{\pm\frac{1}{2},\,\mu}^{\left(s'\right)}\mid_{(-\varphi,\theta,\varphi)}e^{\mp i(\psi+\varphi)/2}=\varPi_{\pm\frac{1}{2},\,\mu}^{\left(s'\right)}\mid_{(\psi,\theta,\varphi)}\in\mathcal{H}_{\mu}^{(s')}$,
$\forall s'\in\left\{ \mathbb{N}_{0}+\frac{1}{2}\right\} $, we can
see that $\left[\left(D_{\mu}^{(s')}\right)^{2}u_{\frac{1}{2}\mu}^{\pm}\right]\mid_{(-\varphi,\theta,\varphi)}$$=\left(\check{\cancel{D}}_{\mathbb{S}^{2}}\right)^{2}Y_{s',\mu}^{\pm}=d_{s'}^{2}Y_{s',\mu}^{\pm}$.
Thus, $\left(D^{(s')}\right)^{2}$ mimics the spectral properties
of $\left(\check{\cancel{D}}_{\mathbb{S}^{2}}\right)^{2}$ for eigenvectors
based on the vectors in $\mathcal{H}_{\mu}^{(s')}$ with biggest negative
and smallest positive $m'\in\underset{\left(2s'+1\right)}{\underbrace{\{-s',-s'+1,...,s'-1,s'\}}}$,
respectively, $\varPi_{-\frac{1}{2},\mu}^{\left(s'\right)}\mid_{(\psi,\theta,\varphi)}$
and $\varPi_{\frac{1}{2},\mu}^{\left(s'\right)}\mid_{(\psi,\theta,\varphi)}$,
as just seen in Proposition 2.3. Thus, the factor $\frac{1}{4}$ in
$d_{s,-\frac{1}{2}}^{2}$ for our $\left(D_{\mu}^{(s')}\right)^{2}$
can be thought as coming from the sphere$.\,\blacksquare$\foreignlanguage{spanish}{ }

$\;$

$\mathbf{Remark\;2.5}$: Certainly, we could define the Dirac operator
directly as $D_{\mu}^{(s)}\doteq-\mathrm{i}\sum_{j=1}^{2}\sigma^{j}X_{j}$,
without any mention to the sphere. But we preferred to do it the other
way around in order to highlight the similitudes and differences between
our spectral triple and that of the sphere (in particular, the factor
$\frac{1}{4}$ in $d_{s,-\frac{1}{2}}^{2}$). We also wanted to show
with this why the Casimir element appears in the area, something which
is not obvious at all, but is generic to Dirac operators (which are
called for by NCG to calculate metric properties) on homogeneous spaces,
as seen in Examples 2.1.1-2. Furthermore, the actual concrete reason
here is that, by the properties of the algebra, $[D,a]_{1,1}$ and
$[D,a]_{1,2}$ \emph{belong again to the algebra}, and this makes
$D^{2}$ appear in the steps for calculating the distance (see Theorem
2.1). Now, the difference is that in NCG one has a whole geometric
machinery to justify why such a Dirac operator must be used to calculate
metric properties (in particular, the reconstruction theorem for the
commutative case, which shows that the information of the classical
metric can be recovered from the Dirac operator), while in LQG one
directly canonically quantizes the classical phase space area functional,
a process which gives us a spectrum that resembles the one of the
Casimir element, but which suffers from the unavoidable obscurity
and black-box-like character of standard canonical quantization. We
believe our approach is more transparent, and that it also gives further
insight (see Remark 2.7-8 later)$.\,\blacksquare$

\selectlanguage{spanish}%
$\;$

Of course, we only saw how Dirac operators of this form just form
the usual spectral triple for the sphere with its corresponding commutative
algebra, $C^{\infty}(\mathbb{S}^{2})$, while what we need here is
a spectral triple with respect to the \emph{non-commutative} algebra
$R_{AQG}\left[\widehat{\mathfrak{su}(2)}\right]$. The obvious option
would be to try to combine \foreignlanguage{english}{$D^{\mathrm{e}}$
on $\mathbb{C}^{2}\otimes\mathcal{H}^{\mathrm{e}}$ with $R_{AQG}\left[\widehat{\mathfrak{su}(2)}\right]$
in some way to obtain a spectral triple. But this won't work because
the $\widehat{\mathrm{E}}_{(j)}(\mathrm{e})$ are \emph{unbounded
operators} on $\mathcal{H}^{\mathrm{e}}$. Thus, we will reduce the
domain of the representation to the \emph{finite dimensional} subspaces
$\check{\mathcal{H}}_{\mu}^{(s)}$ where it's \emph{irreducible}.}

\selectlanguage{english}%
$\;$

$\mathbf{Definition\;2.13}$: The terms in $(\widehat{\mathcal{A}}_{\mu}^{(s)},\check{\mathcal{H}}_{\mu}^{(s)},D_{\mu}^{(s)};\varGamma,J)$
correspond to $\widehat{\mathcal{A}}_{\mu}^{(s)}\doteq$$I_{2\times2}$$R_{AQG}\left[U(\mathfrak{su}(2))\right]_{\mu}^{(s)}$,
and $\varGamma,J$ are the corresponding operators on the $2-$sphere
transferred to our spaces, i.e. $\varGamma\doteq\sigma^{3}I_{\mathcal{H}_{\mu}^{(s)}}$
and, for $\left(\begin{array}{c}
v_{1}\\
v_{2}
\end{array}\right)\in\check{\mathcal{H}}_{\mu}^{(s)}$, $J\left(\begin{array}{c}
v_{1}\\
v_{2}
\end{array}\right)\doteq-T\mathrm{i}\sigma^{2}\overline{\left(\begin{array}{c}
v_{1}\\
v_{2}
\end{array}\right)}=T\left(\begin{array}{c}
-\overline{v}_{2}\\
\overline{v}_{1}
\end{array}\right)$, where $T(f)(g)\doteq f(g^{-1})$$.\,\blacksquare$

$\;$

$\mathbf{Proposition\;2.3}$: The triple $(\widehat{\mathcal{A}}_{\mu}^{(s)},\check{\mathcal{H}}_{\mu}^{(s)},D_{\mu}^{(s)};\varGamma,J)$
is a spectral triple of spectral dimension $0$, on which, \emph{for
elements in} $I_{2\times2}R_{AQG}\left[\widehat{\mathfrak{su}(2)}\right]_{\mu}^{(s)}\doteq\widehat{\mathcal{A}}_{sp.\mu}^{(s)}\subset\widehat{\mathcal{A}}_{\mu}^{(s)}$
(as a Lie algebra only\footnote{For what we do here, only the Lie brackets commutation relations will
be needed, along with the fact that they're realized via the matrix
commutator of the representation spaces (see Remark 2.9 at the end).
Thus, despite having some similitudes, this triple is \emph{not} the
so-called ``fuzzy sphere'', since in the latter the algebra is given
by the usual associative matrix algebra $\mathcal{M}(n,\mathbb{C})$.}), the conditions for the following properties are verified: 
\begin{enumerate}
\item (\emph{twisted}) real structure, 
\item (\emph{un-twisted}) first order, 
\item and $KO-$dimension $2$. 
\end{enumerate}
(This doesn't form a sub-triple in the strict sense, though; on the
other hand, since \emph{only the Lie bracket is needed to prove them},
we could call it a ``\emph{Lie sub-triple}'', since $\widehat{\mathcal{A}}_{sp.\mu}^{(s)}$
is indeed a true Lie sub-algebra of $\widehat{\mathcal{A}}_{\mu}^{(s)}$).$\;$

\emph{Proof}: That $\left[D_{\mu}^{(s)},a\right]$ doesn't vanish
or is ill-defined for $a\in\widehat{\mathcal{A}}_{sp.\mu}^{(s)}$
will be seen below in point 2; both the commutator and $a$ are bounded
because $\check{\mathcal{H}}_{\mu}^{(s)}$ is finite dimensional (and
is also of spectral dimension $0$ for this very reason); furthermore,
$D_{\mu}^{(s)}$ is obviously self-adjoint.

$\;$

1) The only property we need to check is $[a,Jb^{*}J^{-1}]=0,\,\forall a,b\in\mathcal{A}$,
since the others don't depend on the algebra and therefore are identical
to the ones of the $2-$sphere. First, note that the condition is
asking $Jb^{*}J^{-1}=b^{0}$ to be a representation of the \emph{opposite
algebra} $\mathcal{A}^{0}$ (that is, $b^{0}$ is such that $b\in\mathcal{A}$,
but with product $a^{0}b^{0}=(ba)^{0}$, or, in Lie bracket terms,
$[a^{0},b^{0}]=([b,a])^{0}$) that commutes with $\mathcal{A}$. Thus,
given that $J^{-1}\left(\begin{array}{c}
v_{1}\\
v_{2}
\end{array}\right)\doteq T^{-1}\left(\begin{array}{c}
\overline{v}_{2}\\
-\overline{v}_{1}
\end{array}\right)$, $b^{*}=-b$ (by skew-adjointness), and that 
\[
-b^{1,2}f^{1,2}\equiv-X_{b}^{L}(f^{1,2})(g)\doteq-\frac{\mathrm{d}}{\mathrm{d}t}\left[f^{1,2}(ge^{-tb})\right]\mid_{t=0}
\]

\[
=X_{-b}^{L}(f^{1,2})(g),
\]
we get:

\[
b^{0}\left(\begin{array}{c}
f_{1}\\
f_{2}
\end{array}\right)(g)=Jb^{*}J^{-1}\left(\begin{array}{c}
f_{1}\\
f_{2}
\end{array}\right)(g)=J(-b)\left(\begin{array}{c}
\overline{f}_{2}\\
-\overline{f}_{1}
\end{array}\right)(g^{-1})=J\frac{\mathrm{d}}{\mathrm{d}t}\left[\left(\begin{array}{c}
\overline{f}_{2}\\
-\overline{f}_{1}
\end{array}\right)(g^{-1}e^{tb})\right]\mid_{t=0}
\]

\[
=\frac{\mathrm{d}}{\mathrm{d}t}\left[\left(\begin{array}{c}
f_{1}\\
f_{2}
\end{array}\right)(e^{-tb}g)\right]\mid_{t=0}=X_{b}^{R}(\left(\begin{array}{c}
f_{1}\\
f_{2}
\end{array}\right))(g),
\]
which is the \emph{right} representation, and then $[a,b^{0}]=0$,
since $a$ is defined in terms of the \emph{left} representation,
and these two representations \emph{always commute} with each other:

\[
[a,b^{0}]\left(\begin{array}{c}
f_{1}\\
f_{2}
\end{array}\right)(g)=\left[a\frac{\mathrm{d}}{\mathrm{d}t}\left[\left(\begin{array}{c}
f_{1}\\
f_{2}
\end{array}\right)(ge^{tb})\right]\mid_{t=0}-b^{0}\frac{\mathrm{d}}{\mathrm{d}t'}\left[\left(\begin{array}{c}
f_{1}\\
f_{2}
\end{array}\right)(e^{t'a}g)\right]\mid_{t'=0}\right]
\]

\[
=\left[\frac{\mathrm{d}}{\mathrm{d}t'}\frac{\mathrm{d}}{\mathrm{d}t}\left[\left(\begin{array}{c}
f_{1}\\
f_{2}
\end{array}\right)(e^{t'a}ge^{tb})\right]\mid_{t,t'=0}-\frac{\mathrm{d}}{\mathrm{d}t}\frac{\mathrm{d}}{\mathrm{d}t'}\left[\left(\begin{array}{c}
f_{1}\\
f_{2}
\end{array}\right)(e^{t'a}ge^{tb})\right]\mid_{t',t=0}\right]=0,
\]
by commutation of partial derivatives. Note that $JD^{L}J^{-1}=-D^{R}=TD^{L}T^{-1}$,
which means that $J$ is a real structure \emph{only up to the unitary
transformation or ``twist''} $T(\cdot)T^{-1}$, i.e. $JD^{L}J^{-1}\underset{T}{=}D^{L}$.

$\;$

2) $D=-\mathrm{i}(\sigma^{1}X_{1}+\sigma^{2}X_{2})$ and $a=I_{2\times2}\sum_{j=1}^{3}a_{j}X_{j}$.
Then:

\[
[D,a]=-\mathrm{i}\sigma^{1}\sum_{j=1}^{3}a_{j}[X_{1},X_{j}]-\mathrm{i}\sigma^{2}\sum_{j=1}^{3}a_{j}[X_{2},X_{j}]
\]

\[
=-\mathrm{i}\sigma^{1}(a_{2}X_{3}-a_{3}X_{2})-\mathrm{i}\sigma^{2}(-a_{1}X_{3}+a_{3}X_{1}),
\]
and, in this way, since $(a_{2}X_{3}-a_{3}X_{2})$ and $(-a_{1}X_{3}+a_{3}X_{1})$
are in $R_{AQG}\left[\widehat{\mathfrak{su}(2)}\right]_{\mu}^{(s)}$,
they must commute with $\mathcal{A}^{0}$ by point 1), i.e. $[[D,a],b^{0}]=0,\,\forall a,b\in\mathcal{A}$.

$\;$

3) Verbatim as in the case of the $2-$sphere since the operators
involved for the calculation are simply the same$.\,\square$

$\;$

Before continuing, we will need the following examples:

$\;$

$\mathbf{Example\;2.2}$: The following are exemplary of the general
form taken by the hermitian matrices $\mathrm{i}X_{j}$ on the irreducible
spaces $\mathcal{H}_{\mu}^{(s)}$: 
\begin{itemize}
\item $s=1$: 
\end{itemize}
\[
\mathrm{i}X_{1}=\frac{1}{\sqrt{2}}\left(\begin{array}{ccc}
0 & 1 & 0\\
1 & 0 & 1\\
0 & 1 & 0
\end{array}\right),\quad\mathrm{i}X_{2}=\frac{1}{\sqrt{2}}\left(\begin{array}{ccc}
0 & -\mathrm{i} & 0\\
\mathrm{i} & 0 & -\mathrm{i}\\
0 & \mathrm{i} & 0
\end{array}\right),\quad\mathrm{i}X_{3}=\left(\begin{array}{ccc}
1 & 0 & 0\\
0 & 0 & 0\\
0 & 0 & -1
\end{array}\right);
\]

\begin{itemize}
\item $s=\frac{3}{2}$: 
\end{itemize}
\[
\mathrm{i}X_{1}=\frac{1}{2}\left(\begin{array}{cccc}
0 & \sqrt{3} & 0 & 0\\
\sqrt{3} & 0 & 2 & 0\\
0 & 2 & 0 & \sqrt{3}\\
0 & 0 & \sqrt{3} & 0
\end{array}\right),\quad\mathrm{i}X_{2}=\frac{1}{2}\left(\begin{array}{cccc}
0 & -\mathrm{i}\sqrt{3} & 0 & 0\\
\mathrm{i}\sqrt{3} & 0 & -\mathrm{i}2 & 0\\
0 & \mathrm{i}2 & 0 & -\mathrm{i}\sqrt{3}\\
0 & 0 & \mathrm{i}\sqrt{3} & 0
\end{array}\right),
\]

\[
\mathrm{i}X_{3}=\frac{1}{2}\left(\begin{array}{cccc}
3 & 0 & 0 & 0\\
0 & 1 & 0 & 0\\
0 & 0 & -1 & 0\\
0 & 0 & 0 & -3
\end{array}\right);
\]

\begin{itemize}
\item $s\in\frac{1}{2}\mathbb{N}_{0}$:\footnote{We will use the letters $h$, $k$ for this convention of values of
the indexes, and $m$, $\mu$ for the one of Proposition 2.2.} 
\end{itemize}
\[
\mathrm{i}X_{1}^{hk}=\frac{1}{2}(\delta_{h,k+1}+\delta_{h+1,k})\sqrt{(s+1)(h+k-1)-hk},
\]

\[
\mathrm{i}X_{2}^{hk}=\frac{1}{2\mathrm{i}}(\delta_{h,k+1}-\delta_{h+1,k})\sqrt{(s+1)(h+k-1)-hk},
\]

\[
\mathrm{i}X_{3}^{hk}=(s+1-k)\delta_{hk},\;1\leq h,k\leq(2s+1).\,\blacksquare
\]

$\;$

$\mathbf{Theorem\;2.1}$: The \emph{noncommutative distance} for the
previous triple\emph{ for elements in} $\widehat{\mathcal{A}}_{sp.\mu}^{(s)}$,
between ``the origin'' $\Omega_{-}$ and the pure state $\Omega_{+}$
at the maximum possible distance from it, is 
\[
\cancel{d}\,(\Omega_{+},\Omega_{-})=d_{s}^{-1},
\]
where $d_{s}=d_{s,-\frac{1}{2}}$ for $2s+1$ even, and $d_{s}=d_{s,0}$
for $2s+1$ odd.

$\;$

\emph{Proof}: the operator norm of $T$ is defined as $\parallel T\parallel^{2}\doteq\underset{v\in\mathcal{H}}{\mathrm{sup}\,}\left\{ (v,T^{*}Tv)_{\mathcal{H}}\diagup\parallel v\parallel=1\right\} $.
We compute:

\[
[D,a]^{*}[D,a]=\left[\mathrm{i}\sigma^{1}(a_{2}X_{3}-a_{3}X_{2})+\mathrm{i}\sigma^{2}(-a_{1}X_{3}+a_{3}X_{1})\right]\left[\mathrm{i}\sigma^{1}(a_{2}X_{3}-a_{3}X_{2})+\mathrm{i}\sigma^{2}(-a_{1}X_{3}+a_{3}X_{1})\right]
\]

\[
=-I_{2\times2}(a_{2}X_{3}-a_{3}X_{2})^{2}-\sigma^{2}\sigma^{1}(-a_{1}X_{3}+a_{3}X_{1})(a_{2}X_{3}-a_{3}X_{2})-\sigma^{1}\sigma^{2}(a_{2}X_{3}-a_{3}X_{2})(-a_{1}X_{3}+a_{3}X_{1})
\]

\[
-I_{2\times2}(-a_{1}X_{3}+a_{3}X_{1})^{2}
\]

\[
=-I_{2\times2}(a_{2}^{2}X_{3}^{2}-a_{2}a_{3}\{X_{3},X_{2}\}+a_{3}^{2}X_{2}^{2})+\sigma^{1}\sigma^{2}[(-a_{1}X_{3}+a_{3}X_{1}),(a_{2}X_{3}-a_{3}X_{2})]
\]

\[
-I_{2\times2}(a_{1}^{2}X_{3}^{2}-a_{1}a_{3}\{X_{3},X_{2}\}+a_{3}^{2}X_{1}^{2})
\]

\[
=a_{3}^{2}D^{2}+I_{2\times2}a_{3}(a_{1}+a_{2})\{X_{2},X_{3}\}+I_{2\times2}(a_{1}^{2}+a_{2}^{2})(-X_{3}^{2})-\sigma^{3}a_{3}(a_{1}\mathrm{i}X_{1}+a_{2}\mathrm{i}X_{2}).
\]
Consider, now, any $v=\omega\otimes V\in\mathbb{C}^{2}\otimes\mathcal{H}$
such that $\parallel\omega\otimes V\parallel=1$ (recall that $(\omega\otimes V,\omega\otimes V)_{\mathbb{C}^{2}\otimes\mathcal{H}}=(\omega,\omega)_{\mathbb{C}^{2}}(V,V)_{\mathcal{H}}$).
We would, then, get something of the form:

\[
(v,[D,a]^{*}[D,a]v)_{\mathbb{C}^{2}\otimes\mathcal{H}}=a_{3}^{2}D^{2}v+\parallel\omega\parallel^{2}\sum_{\alpha=0,3}(V,M_{\alpha}V)_{\mathcal{H}}+\sigma(\omega)\sum_{\alpha=1,2}(V,M_{\alpha}V)_{\mathcal{H}},
\]
with

\[
M_{0}=a_{3}(a_{1}+a_{2})\{X_{2},X_{3}\},\quad M_{1}=-a_{3}a_{1}\mathrm{i}X_{1},\quad M_{2}=-a_{3}a_{2}\mathrm{i}X_{2},\quad M_{3}=(a_{1}^{2}+a_{2}^{2})(-X_{3}^{2}),
\]

\[
(V,M_{\alpha}V)_{\mathcal{H}}=\sum_{h,k=1}^{2s+1}M_{\alpha}^{hk}V^{h}\overline{V}^{k},\quad\sigma(\omega)=(\omega,\sigma^{3}\omega)_{\mathbb{C}^{2}}=\mid\omega_{1}\mid^{2}-\mid\omega_{2}\mid^{2}.
\]
Now, we claim that, \emph{for each} $a$, \emph{we have} 
\[
\parallel[D,a]\parallel^{2}-a_{3}^{2}d_{s}^{2}\doteq Q(a)^{2}\geq0.
\]
Thus, we get\footnote{the square power is just for emphasis regarding the positivity of
the term; $Q(a)$ is, of course, \emph{real}.}:

\[
\parallel[D,a]\parallel=\sqrt{a_{3}^{2}d_{s}^{2}+Q(a)^{2}}.
\]
It remains to prove the claim. The problematic terms are those of
$M_{0}$, $M_{1}$, and $M_{2}$, since, in them, the components $a_{1}$,
$a_{2}$, $a_{3}$, appear \emph{unsquared} so that their sign could
make $\parallel[D,a]\parallel^{2}-a_{3}^{2}d_{s}^{2}$ \emph{negative},
because the growth of the only remaining term, $(V,M_{3}V)_{\mathcal{H}}$
---which is always strictly positive--- is \emph{bounded} from above
by the norm of $\mathrm{i}X_{3}$ (which equals $\parallel\mathrm{i}X_{3}\parallel_{\mathcal{B}(\mathcal{H})}=s$),
that is: $\parallel\mathrm{i}X_{3}V\parallel_{\mathcal{H}}\leq\parallel\mathrm{i}X_{3}\parallel_{\mathcal{B}(\mathcal{H})}\parallel V\parallel_{\mathcal{H}}$,
$\forall V\in\mathcal{H}$.

Consider the normalized eigenvector $V_{\left(3\right)}=(0,..,\underset{m=-\frac{1}{2};0}{\underbrace{1}},...,0)$
of $-X_{3}^{2}$ with the biggest value, $d_{s}^{2}$, for $d_{s,m}^{2}$
(then, for even $2s+1$ we get $m_{3}=-\frac{1}{2}$ and so $d_{s}^{2}=s(s+1)+\frac{1}{4}$,
and for odd $2s+1$ we get $m_{3}=0$ and so $d_{s}^{2}=s(s+1)$)
where $\omega_{\left(3\right)}=(1,0)$. We then get ($v_{\left(3\right)}=\omega_{\left(3\right)}\otimes V_{\left(3\right)}$):

\[
(v_{\left(3\right)},[D,a]^{*}[D,a]v_{\left(3\right)})_{\mathbb{C}^{2}\otimes\mathcal{H}}-a_{3}^{2}d_{s}^{2}=m_{3}^{2}(a_{1}^{2}+a_{2}^{2})\geq0,
\]
since\footnote{$\{X_{2},X_{3}\}^{hk}=(\delta_{h,k+1}-\delta_{h+1,k})(2\mathrm{i})(2s+2-h-k)\sqrt{(s+1)(h+k-1)-hk}.$}
$M_{\alpha}^{-\frac{1}{2},\,-\frac{1}{2}}=0$, $M_{\alpha}^{00}=0$,
$\forall\alpha\neq3$. Thus, given that \emph{there's at least one
$v_{\left(3\right)}\in\mathbb{C}^{2}\otimes\mathcal{H}$ satisfying}
\[
(v_{\left(3\right)},[D,a]^{*}[D,a]v_{\left(3\right)})_{\mathbb{C}^{2}\otimes\mathcal{H}}\geq a_{3}^{2}d_{s}^{2}\geq a_{3}^{2}d_{s,m}^{2},
\]
and that, by definition, 
\[
\parallel[D,a]\parallel^{2}\geq(v,[D,a]^{*}[D,a]v)_{\mathbb{C}^{2}\otimes\mathcal{H}},\,\forall v\in\mathbb{C}^{2}\otimes\mathcal{H},
\]
we must \emph{necessarily} have

\[
\parallel[D,a]\parallel^{2}\geq(v_{\left(3\right)},[D,a]^{*}[D,a]v_{\left(3\right)})_{\mathbb{C}^{2}\otimes\mathcal{H}}\geq a_{3}^{2}d_{s}^{2}\geq a_{3}^{2}d_{s,m}^{2},
\]
which means that

\[
\parallel[D,a]\parallel^{2}-a_{3}^{2}d_{s}^{2}\doteq Q(a)^{2}\geq0.
\]
Consider now the density matrices $\rho^{+}$, $\rho^{-}$(which represent
\emph{pure} states in $\widehat{\mathcal{A}}_{\mu}^{(s)}$ in this
case) given by: 
\begin{itemize}
\item $n=2s+1$ even: $\rho_{ij}^{+}=1$ for $i=j=\frac{n}{2}$, and $\rho_{ij}^{+}=0$
otherwise; $\rho_{ij}^{-}=1$ for $i=j=\frac{n}{2}+1$, and $\rho_{ij}^{-}=0$
otherwise. 
\item $n=2s+1$ odd: $\rho_{ij}^{+}=1$ for $i=j=\frac{n-1}{2}$ and $\rho_{ij}^{+}=0$
otherwise; $\rho_{ij}^{-}=1$ for $i=j=\frac{n-1}{2}+1$ and $\rho_{ij}^{-}=0$
otherwise. 
\end{itemize}
Then, for an arbitrary $a\in\widehat{\mathcal{A}}_{sp.\mu}^{(s)}$,
we get: 
\begin{itemize}
\item even case: $\Omega_{\rho^{+}}(a)\doteq$$\mathrm{tr}\,(\rho^{+}a)=\frac{1}{2}\mathrm{i}a_{3}$
and $\Omega_{\rho^{-}}(a)\doteq$$\mathrm{tr}\,(\rho^{-}a)=-\frac{1}{2}\mathrm{i}a_{3}$; 
\item odd case: $\Omega_{\rho^{+}}(a)=$$\mathrm{tr}\,(\rho^{+}a)=\mathrm{i}a_{3}$
and $\Omega_{\rho^{-}}(a)=$$\mathrm{tr}\,(\rho^{-}a)=0$. 
\end{itemize}
Thus, the noncommutative distance is:

\[
\cancel{d}\,(\Omega_{\rho^{+}},\Omega_{\rho^{-}})=\mathrm{sup}\,\left\{ \mid a_{3}\mid,\,\forall a\in\mathcal{A}\diagup\parallel[D,a]\parallel\leq1\right\} .
\]
Going back to $\parallel[D,a]\parallel$: since it's a sum of \emph{positive}
terms, $\parallel[D,a]\parallel=\sqrt{a_{3}^{2}d_{s}^{2}+Q(a)^{2}}$,
and $a_{3}^{2}d_{s}^{2}$ \emph{only depends on} $a_{3}$, it's therefore
clear that the second term should go away if we want to maximize $\mid a_{3}\mid$
under the constraint $\parallel[D,a]\parallel\leq1$ (otherwise, when
varying $a_{1}$, $a_{2}$, the other positive term will diminish
the part of $1$ that goes to $a_{3}^{2}d_{s}^{2}$, so to speak),
so our only option is $a_{\mathrm{sup.}}=(0,0,a_{3})$, where we get
$Q(a_{\mathrm{sup.}})^{2}=0$. In this way, $\parallel[D,a_{\mathrm{sup.}}]\parallel=\sqrt{a_{3}^{2}d_{s}^{2}}=\mid a_{3}\mid d_{s}=1$,
i.e. $\mid a_{3}\mid=d_{s}^{-1}$. Therefore, $\cancel{d}=d_{s}^{-1}$$.\,\square$

$\;$

$\mathbf{Remark\;2.6}$: Note that the actual \emph{physical} distance
will be $\cancel{d}=d_{s}$ because the triple we have is for the
algebra, while our initial variables were the quantization of momenta
on the \emph{dual} algebra (Lemma 2.2), which means that the physical
metric is the \emph{inverse} (or dual) of the metric on the algebra$.\,\blacksquare$

$\;$

$\mathbf{Remark\;2.7}$:\emph{ }The obtained values for the area\emph{
are precisely the only values allowed by the area operator from LQG
for such a punctured surface}\footnote{All of this, of course, also applies to the surfaces dual to \emph{each}
of the edges $\mathrm{e}$ in the graph $\alpha$, and one can also
consider other graphs as well. Since the representation is carried
out on the tensor product space, this means that the area values for
the surfaces corresponding to \emph{different} edges are just \emph{added}
for obtaining the total area for the ``union'' of those surfaces,
which now will have \emph{two punctures}; again, this is exactly the
result that one gets from the area operator in standard LQG.}\emph{,} since we can make $\left(\frac{1}{4\rho_{\,_{\mathbb{S}^{2}}}^{2}}\right)$
\emph{as small as we want} by just letting $\rho_{\,_{\mathbb{S}^{2}}}$
get bigger. Nevertheless, \emph{our approach makes a further prediction}:
$d_{s}$ are just the biggest distance values, but by choosing (recall
that $n=2s+1$) $\rho_{ij}^{+}=1,\,i=j=\frac{n}{2}-h$ ($0\leq h\leq\frac{n}{2}-1$),
$\rho_{ij}^{-}=1,\,i=j=\frac{n}{2}+k$ ($1\leq k\leq\frac{n}{2}$),
for the even case, and $\rho_{ij}^{+}=1,\,i=j=\frac{n-1}{2}-h$, ($0\leq h\leq\frac{n-1}{2}-1$),
$\rho_{ij}^{-}=1,\,i=j=\frac{n-1}{2}+k$ ($1\leq k\leq\frac{n+1}{2}$),
for the odd case, \emph{we get smaller values}:

\[
\cancel{d}_{h,k}=\frac{d_{s}}{h+k}\,.
\]
The physical relevance and consequences of this difference with respect
to the prediction of LQG remains a topic for further research.$\,\blacksquare$

$\;$

$\mathbf{Remark\;2.8}$: The triple has \emph{spectral dimension}
$0$ and $KO-$dimension $2$. This seems to be a clear case of \emph{spectral
dimensional drop} (from $2$, in the classical case\footnote{We are \emph{not} referring to the $2-$sphere of Example 2.1.2-2
(which was just a mathematical auxiliary for the construction of $D$),
but to some arbitrary surface in spacetime, whose coordinate algebra
was related to the $\mathrm{E}(S,\boldsymbol{k})$ variables from
GR that were quantized here. }, to $0$ here) upon quantization\footnote{The striping out of the continuum played a key role on this, since
it transformed the \emph{infinite} dimensional (real) algebra of continuous
functions of the classical case into a \emph{finite} dimensional one,
thus opening the possibility of irreducible representations on finite
dimensional spaces. } for the surface. The interpretation that we make of the fact that
the irreducible geometries are $0-$dimensional is the following:
recall, from Example 1.5-6, that the spectral triples that describe
the geometry of a set with a \emph{finite} number of points, and the
Euclidean distances among them, are spectral $0-$dimensional, too.
There's an important difference, though, with respect to the classical
points, since here what's coming as a noncommutative \textit{distance}
should actually be physically interpreted as an \emph{area}. This
means that the ``points'' must be some sort of \emph{irreducible}
``string-like'' objects at the physical level. \emph{Thus, this
approach offers new and more detailed insight into the nature of area
in QG}$.\,\blacksquare$

$\;$

$\mathbf{Remark\;2.9}$: The reason for working on the triple based
on the ``ambient'' given by $R_{AQG}\left[U(\mathfrak{su}(2))\right]$
lies in the fact that the matrix product of the representation space
doesn't close on the Lie algebra $R_{AQG}\left[\widehat{\mathfrak{su}(2)}\right]$,
whereas the Lie bracket in terms of that product does. On the other
hand, on physical grounds, only the use of elements in $\mathfrak{su}(2)$
is justified, and that's why only that sub-space of elements was considered
when verifying the properties and in the calculation of the distance.
That is, we make the hypothesis that the latter \emph{lower bound}
on the distance for the triple of Definition 2.13 (or, the \emph{true
distance} of what we called a ``Lie sub-triple'') \emph{is the actual
physical distance}$.\,\blacksquare$

$\;$

\subsection*{{\normalsize{}{}3. Conclusions}}

$\;$

In NCG, what determines the structure of spacetime (in particular,
if it's a classical differentiable manifold or not) is its algebra
$\mathcal{U}_{st}$. Inspired by a relationalist conception of spacetime,
we made a detailed analysis of the relation between this algebra and
the phase space algebra of the Gravitational Field. We proposed an
approach which, using mathematical tools from non-commutative geometry
(NCG) à  la Connes and the Topos approach to quantum theory (Isham-Doering),
sheds new light on the obscure issue of the space(-time) picture in
(canonical) quantum gravity. We then applied our scheme to the particular
algebra of Algebraic Quantum Gravity, a cousin of LQG best suited
to our purposes. In this context, we obtained a novel way of deducing
the quantization of the possible values for the area of a surface,
which came expressed in a \textit{purely combinatorial} way (this
being a long sought thing in the context of canonical LQG \cite{key-15,key-16}).
We got the same values than those of LQG, but \textit{also additional,
smaller values}. Last but not least, \textit{at no point of our derivation
}(in the quantized theory) \textit{did we use the classical continuum
manifold}, thus avoiding the possibility of our assumptions contradicting
our conclusions, as well as obtaining a physically clear picture of
quantum space in return. In LQG it has often been repeated that ``edges
of a graph carry quanta of area''. This however is only \textit{indirectly}
and \textit{partially} hinted in LQG by the quantized phase space
functionals appearing there, but not \textit{explicitly} since spatial
surfaces and regions (which are the very things that carry areas and
volumes) can only be identified relationally\textit{ once a solution
has been chosen}, and not at the level of the whole phase space, where
they're devoid of physical meaning. Instead, this claim becomes fully
realized and rigorously established under our approach. We directly
obtain a well defined non-commutaive space (in the sense of Connes'
NCG, that is, a so-called \emph{spectral triple}) whose non-commutative
metric geometry (characterized by Dirac-like operators) is quantized
and related to graphs in the same way as in LQG. These spaces can
be identified with the physical space that relationaly arises from
the cannonically quantized Gravitational Field (quantum relational
space being another long-sought entity in the LQG approach \cite{key-9}).

$\;$

\end{document}